\documentclass[useAMS,usenatbib]{mn2e}

\usepackage{commonPackages}
\usepackage{mathsCommands}
\usepackage{bibCommands}

\title[Accreting millisecond pulsar magnetospheres]{Simulations of the magnetospheres of accreting \\ millisecond pulsars}

\author[Kyle Parfrey, Anatoly Spitkovsky, and Andrei M. Beloborodov]
{Kyle Parfrey$^{1}$\thanks{E-mail: kparfrey@lbl.gov}\thanks{Einstein Fellow}, Anatoly Spitkovsky$^2$, and Andrei M. Beloborodov$^3$\\
$^{1}$Lawrence Berkeley National Laboratory, 1 Cyclotron Road, Berkeley, CA 94720, USA\\
$^{2}$Department of Astrophysical Sciences, Princeton University, Peyton Hall, Princeton, NJ 08544, USA\\
$^{3}$Department of Physics, Columbia University, 538 West 120th Street, New York, NY 10027, USA}

\begin{document}

\maketitle

\begin{abstract}
  Accreting pulsars power relativistic jets, and display a complex spin phenomenology.  These behaviours may be closely related to the large-scale configuration of the star's magnetic field, shaped by its interaction with the surrounding accretion disc. Here we present the first relativistic simulations of the interaction of a pulsar magnetosphere with an accretion flow. Our axisymmetric simulations treat the magnetospheric, or coronal, regions using a resistive extension of force-free electrodynamics. The magnetic field is also evolved inside the disc, which is a defined volume with a specified velocity field and conductivity profile, found using an $\alpha$-disc model.  We study a range of disc $\alpha$-parameters, thicknesses, magnetic Prandtl numbers, and inner truncation radii. We find that a large fraction of the magnetic flux in the pulsar's closed zone is opened by the intrusion of the disc, leading to an enhancement of the power extracted by the pulsar wind and the spin-down torque applied to the pulsar. In our simulations, most of the spin-down contribution to the stellar torque acts on open field lines. The efficiency of field-line opening is high in the simulations' long-term quasi-steady states, which implies that a millisecond pulsar's electromagnetic wind could be strong enough to power the observed neutron-star radio jets, and may significantly affect the pulsar's spin evolution.
\end{abstract} 

\begin{keywords}
  pulsars: general -- stars: jets -- stars: neutron -- X-rays: binaries -- accretion, accretion discs 
  -- magnetic fields
\end{keywords}

\section{Introduction} 

Many characteristics of neutron stars in low-mass X-ray binaries are attributed to the interaction of the star's magnetic field with a flow of dense accreting plasma, supplied by the companion via Roche-lobe overflow. These features include magnetically channeled accretion leading to pulsed X-ray emission \citep{Gnedin:1973aa, Wijnands:1998aa}, relativistic radio jets \citep{Fomalont:2001ab, Fender:2004aa, Migliari:2006aa}, and spin-up of the neutron star to nearly kiloHertz frequencies \citep{Chakrabarty:2003aa}.

As much of the energy and angular momentum transfer is mediated by magnetic fields, the form taken by the global magnetic geometry has far-reaching consequences. Theories of magnetic star--disc systems can be roughly divided into those in which the field remains approximately dipolar, leading to extensive magnetic coupling between the stellar surface and the disc \citep{Ghosh:1977aa, Ghosh:1978aa, Ghosh:1979aa, Wang:1995aa, Rappaport:2004aa}, those in which the field is excluded from the disc by surface currents \citep{Aly:1980aa}, and models in which most of the stellar flux that would enter the disc is opened, breaking the star--disc connection \citep{Camenzind:1990aa,Shu:1994aa,Lovelace:1995aa}. This flux opening is a direct result of the angular-velocity mismatch that generally exists between the footpoint of a field line on the star and where it enters the disc \citep{Lynden-Bell:1994aa,Uzdensky:2002ac}, and may be responsible for observed flaring activity via magnetic reconnection in the ensuing current sheet \citep{Aly:1990aa}.

Stressed (i.e. twisted) field lines which link the star and the disc allow angular momentum to be exchanged between them, acting to increase the star's rotation rate when the field line enters the disc inside the corotation radius, $\rco = \left( G M / \Omega^2\right)^{1/3}$, where $M$ and $\Omega$ are the star's mass and spin angular frequency, and applying a braking torque on the star if it enters beyond this point. How field lines respond to these stresses depends on the effective magnetic diffusivity in the disc, provided by the magnetorotational turbulence. When the diffusivity is high, field lines move outward through the disc \citep{Bardou:1996aa}, reducing the torque that they apply to the star \citep{Agapitou:2000aa}. If the diffusivity is low, the stressed field lines are opened to infinity, and the star--disc torques they carried are replaced by an outflow of angular momentum from both the star and the disc material, carried by their individual rotationally driven magnetohydrodynamic (MHD) winds \citep{Lovelace:1995aa}.

The strength of the star's MHD wind is determined by its rotation rate, magnetic field strength, and the outflowing inertia threaded by the field. In the non-relativistic limit this last component is the wind's mass loading, and so for systems like accreting protostars the open field lines must be continuously supplied with sufficient matter if they are to exert a significant torque on the star \citep[e.g.][]{Ferreira:2000aa}. On the other hand, millisecond pulsars have spin frequencies of several hundred Hertz and surface magnetic field strengths of roughly $10^8$~G, and so the torque applied by the relativistic pulsar wind, in which the inertia in question is the effective inertia of the electromagnetic field itself, can be significant.

In \citet[henceforth Paper I]{Parfrey:2016aa} we described estimates of the spin-down torque due to the relativistic pulsar wind, when the amount of open stellar flux has been increased by interaction with the disc via the process described above; see also Section~\ref{sec:model}. For rapidly rotating pulsars this enhanced wind torque can be of the order of the estimated spin-up torque from accreting material, and for efficient field opening it can be strong enough to enforce spin equilibrium at frequencies less than 1 kHz for reasonable stellar magnetic field strengths and accretion rates. Estimates of the energy flux on the open field lines suggest that the enhanced pulsar wind could be responsible for even the most extreme relativistic jets from neutron star systems, if the electromagnetic outflow is collimated by an external agent, for example an MHD wind from the outer accretion disc. In this work we present the results of simulations which directly test the model of Paper I, in particular the efficiency of flux opening and the relative importance of open and disc-coupling flux in determining the stellar spin-down torque.

MHD simulations of the star--disc system allow the global magnetic field and gas flow to be evolved self-consistently, including the reaction of the stellar field on the disc. Simulations of this kind have displayed a wide range of behaviours in different regimes, including ejections of plasmoids and magnetic braking of the inner disc \citep{Hayashi:1996aa,Miller:1997aa}, quasi-periodic opening and reconnection of flux coupling the star and disc \citep{Goodson:1997aa,Goodson:1999aa,Zanni:2013aa}, funnel flows of accreting gas \citep{Romanova:2002aa,Bessolaz:2008aa}, the launching of powerful jets by magnetic towers \citep{Kato:2004aa,Romanova:2009aa},  and `propellering' of gas against gravity by rapid rotators \citep{Romanova:2004aa}. Coupling of the stellar field to the disc out to large radii was observed when the disc magnetic diffusivity was high \citep{Zanni:2009aa}. Simulations have been performed in which the disc effective viscosity and magnetic diffusivity were determined by the self-consistent evolution of the magnetorotational instability \citep[MRI;][]{Romanova:2011aa}. As an alternative to evolving both the disc and its corona in the same domain, some simulations have incorporated the disc as a boundary condition \citep[e.g.][]{Fendt:1999aa,Fendt:2000aa}.

Here we take a different, in some ways simplified, approach. We evolve the magnetically dominated region, the magnetosphere or corona, using relativistic force-free electrodynamics, the high-magnetisation limit of MHD \citep{Uchida:1997aa,Gruzinov:1999aa,Komissarov:2002aa}, supplemented with small resistive corrections. The electromagnetic fields are also evolved inside the disc volume, which is modeled as a conducting fluid with an imposed velocity field and conductivity profile. This method has two principal advantages.

Firstly, it allows the magnetosphere to be evolved with zero mass loading, in which all the inertia comes from the electromagnetic field itself. For open field lines, this gives the minimal possible magnetic stresses -- any spin-down torque communicated by the open flux must be at least as large as in the force-free case, provided enough charge carriers are present to supply the currents. The coronae of fairly thin or slim discs in particular are expected to have very low mass density due to the small disc scale-height, and so to be highly magnetically dominated. This regime is problematic for simulations using the full MHD equations, which require a density floor for numerical stability. Secondly, because the disc is imposed artificially as part of the problem specificiation, this approach permits a wide-ranging investigation of how the star--disc magnetosphere depends on various disc properties, such as thickness, conductivity, and the radial accretion velocity. 

On the other hand, our simplified method has several limitations. In particular, the accretion disc does not evolve self-consistently, and so we neglect, for example, the effect of angular momentum extraction into a disc wind by open field lines, which may increase the radial accretion speed above the $\alpha$-disc estimates. Rather than aiming to provide a conclusive overall picture of these highly complex systems, we have focused on two principal issues: the efficiency of magnetic flux opening, and the contribution of open flux to torques and wind powers.

In our simulations the only magnetic field is the stellar dipole. In effect, we assume that the disc's magnetic field is present exclusively at small scales, where it generates MRI turbulence and hence the accretion velocity and effective disc conductivity that we specify \citep[e.g.][]{Eardley:1975aa,Balbus:1991aa}. In reality the disc's field may well mediate the coupling between stellar field and disc through reconnection near the disc's surface; this will not appreciably change any of our conclusions. We defer an investigation of the effect of large-scale magnetic structures in the accretion flow to future work.

We begin by briefly summarizing in Section~\ref{sec:model} the enhanced-pulsar-wind model of Paper I. The numerical method and simulation procedure is described in detail in Section~\ref{sec:method}. Our principal simulations, in which the accretion-flow properties are found using an $\alpha$-disc model, are presented in Section~\ref{sec:simsAlpha}, while a large set of further-idealized simulations are analyzed in Section~\ref{sec:simsGrid}. In Section~\ref{sec:comparison} we compare to non-relativistic simulations of accretion onto magnetized stars. We discuss our results and describe our conclusions in Section~\ref{sec:conclusions}.


\section{Enhanced pulsar wind model}
\label{sec:model}

The magnetosphere of an isolated pulsar has a region of approximately dipolar closed field lines within the light cylinder which is defined by its rotation,
\begin{equation}
  \rlc = \frac{ c }{ \Omega },
\end{equation}
and beyond which all field lines are open; $\Omega$ is the star's spin angular frequency. The pulsar wind power $L_0$ and the torque exerted by the magnetosphere on the star $N_0$ are given by 
\begin{equation}
  \Lo = - \No \Omega = \mu^2 \frac{\Omega^4}{c^3} \approx \frac{2}{3c}\Omega^2\ffopeno^2,
\end{equation}
where $\mu$ is the star's magnetic moment and $\ffopeno$ is the amount of open flux \citep{Goldreich:1969aa,Contopoulos:2005aa,Gruzinov:2005aa}; throughout we use a subscripted `0' to indicate quantities referring to an isolated pulsar.  In Paper I we described a simple analytical model in which a surrounding disc, inside the light cylinder, opens some of the closed flux via the mechanism described above, causing the amount of open flux $\ffopen$ to increase,
\begin{equation}
  \ffopen = \zeta \frac{\rlc}{\rmag} \ffopeno \,,
  \label{eq:openFluxModel}
\end{equation}
where we have assumed a dipolar dependence of the poloidal flux function $\ff$ on radius in the closed zone: $\ff \propto 1/r$ at the equator; here the magnetospheric radius $\rmag$ is the inner truncation radius of the disc, and $\zeta$ parameterizes the flux-opening efficiency. The poloidal flux function is defined as
\begin{equation}
  \ff(r,\theta) = \int_{\phi'=0}^{2\pi} \int_{\theta'=0}^\theta B_r \, r^2 \sin\theta' \rm{d}\theta'\rm{d}\phi'.
\end{equation}
When we give numerical values of $\ff$, we normalize such that $\ff(\rstar, \pi/2) = 1$, where $\rstar$ is the stellar radius [i.e. the dipole field is given by $\ff = \sin^2\theta\, (\rstar/r)$].

Because the spin-down power and torque simply scale quadratically with the open flux, straightforward estimates can be made,
\begin{align}
  \Nopen &= \zeta^2 \left( \frac{\rlc}{\rmag}\right)^2 N_0, \label{eq:torqueModel} \\ 
  \Lj &= \zeta^2 \left( \frac{\rlc}{\rmag}\right)^2 L_0, \label{eq:jetModel}
\end{align}
where we have associated the spin-down power on the open field lines with the system's jet power, $\Lj$. If the disc conductivity is high we expect most field lines to be open and $\Nopen$ to be a good estimate of the total spin-down torque on the star --- this proposal is tested in Section~\ref{sec:simsGrid}. 

All one now needs to make estimates of the strength of the open-flux torque and electromagnetic jet power is a model for the disc truncation radius. In Paper I we used 
\begin{equation}
  \rmag = \xi \, \ra \,,
\end{equation}
where $\ra = (\mu^4/2GM\mdot^2)^{1/7}$ is the standard estimate of the Alfv\'{e}n radius assuming spherical accretion. MHD simulations have usually found $\xi \sim 0.5$--$0.7$ \citep{Long:2005aa,Bessolaz:2008aa,Zanni:2013aa}. A steady state of the accretion flow and magnetosphere is then specified by the model parameters $\zeta$ and $\xi$. The model provides several simple relationships among the quantities of interest; for example, the jet power should scale roughly as $\Lj \propto \mdot^{4/7}$. If the total spin-up torque is dominated by the accretion torque, $\Nacc = \mdot \sqrt{GM\rmag}$, spin equilibrium (zero torque on the star) occurs when $\rmag/\rlc = 2^{-1/2} \xi^{7/2}/\zeta^2$. See Paper I for further discussion of the model and the relevant observations.

It will be useful to have an estimate of the strength of the spin-up torque exerted by accreting matter, $\Nacc$, in units of the isolated pulsar spin-down torque, as the accretion funnel and associated sub-Keplerian gas flow are not included in our simulations. Using the above expressions for $\rmag$, $\ra$, $\Nacc$, and $N_0$, one finds
\begin{equation}
  \frac{\Nacc}{N_0} = \frac{\xi^{7/2}}{\sqrt{2}} \left( \frac{\rlc}{\rmag} \right)^3\,.
  \label{eq:Nacc}
\end{equation}


\section{Numerical Method}
\label{sec:method}

We divide the space outside the star into two regions: the disc and the magnetosphere. We solve Maxwell's equations,
\begin{equation}
  \frac{1}{c}\frac{\dd\B}{\dd t} = - \curlE, \;\;\;\;
  \frac{1}{c}\frac{\dd\E}{\dd t} = \curlB - \frac{4\pi}{c}\,\J,
\end{equation}
in the complete domain using the pseudospectral \textsc{phaedra} code \citep{Parfrey:2012ab}, subject to each region's appropriate current density $\J$: in the disc we set $\J = \J_{\rm disc}$, while in the magnetosphere $\J = \J_{\rm FFE}$. 

We take the magnetosphere to be magnetically dominated, and therefore base our approach on force-free electrodynamics, the high-magnetic-field (alternatively, vanishing-matter-inertia) limit of MHD. We use a resistive extension of the force-free current density,
\begin{align}
  \frac{4\pi}{c}\, \J_{\rm FFE}& = \ls \B\cdot\curlB - \E\cdot\curlE + \gamma\E\cdot\B \rs \frac{\B}{(1+\gamma\eta)B^2} \nonumber \\
                                 & \quad + \divE\,\, \frac{\E\times\B}{B^2 + \tilde{E}^2},
  \label{eq:Jffe}
\end{align}
where the denominator in the $\E\times\B$ drift current is defined via \citep{Gruzinov:2008aa,Li:2012aa},
\begin{align}
  \tilde{E}^2& = \tilde{B}^2 + E^2 - B^2, \nonumber \\
  \tilde{B}^2& = \frac{1}{2} \ls B^2 - E^2 + \sqrt{\lp B^2 - E^2\rp^2 + 4 \lp \E\cdot\B \rp^2} \rs .
\end{align}

The current along the magnetic field supports Alfv\'{e}n waves, while fast magnetosonic waves do not carry current. Slow magnetosonic and sound waves are not present, as the gas pressure is assumed to be negligible in comparison to the magnetic pressure. The current perpendicular to $\E$ and $\B$ is due to advection of charge at the $\E\times\B$ drift velocity.

Here $\eta$ is the resistivity and $\gamma$ is a driving rate: $\E\cdot\B$ is driven to $\eta \J\cdot\B$ over a time $1/\gamma$. This form of resistivity only dissipates currents along the magnetic field, so Alfv\'{e}n waves are dissipated but fast magnetosonic waves are unaffected. As $\eta \rightarrow 0$, ideal force-free electrodynamics is recovered, in which $\E\cdot\B = 0$ and $\rhoe \E + \J\times\B/c = 0$, where $\rhoe = \divE/4\pi$ is the charge density. 

We use a very small $\eta$ in the magnetosphere, so that relaxation of twisted field lines, and the triggering of reconnection in current sheets, is due to physically modeled dissipative electric fields rather than the numerical dissipation intrinsic to the simulations. The resistivity is modeled with a constant background and a current-dependent term which only becomes appreciable in current sheets,
\begin{equation}
  \eta = \eta_{\rm bg} + \eta_J \frac{|\J_{\rm ideal}\cdot\B|}{B^2 + E^2},
\end{equation}
and a maximum value of $\eta_{\rm max}$ is enforced; $\J_{\rm ideal}$ is the usual ideal force-free current, i.e equation~(\ref{eq:Jffe}) with $\eta$ and $\gamma$ set to zero. We set $\eta_{\rm bg}/4\pi = 5 \times 10^{-6} \, \rstar/c$, $\eta_J/4\pi = 2 \times 10^{-5} \, \rstar/c$, and $\eta_{\rm max} = 10 \eta_{\rm bg}$. The magnetospheric background conductivity, $4\pi\sigma_{\rm bg} = 2\times 10^5 c/\rstar$, is about two orders of magnitude larger than our highest disc conductivity. The driving rate $\gamma$ is chosen at each point to be the instantaneous maximum numerically stable value. 

A significant problem when evolving the force-free equations, ideal or resistive, is the spontaneous appearance of regions where $E > B$, implying the breakdown of the assumptions on which the system of equations is based. This usually occurs during current-sheet formation, and is accompanied by numerical instability. Here we use a `current-sheet capturing' method to find these pathological regions between grid points, and smoothly remove electric field from the immediate vicinity such that current sheets do not collapse to the grid scale and stability is maintained. The resistive force-free equations and current sheet capturing method will be described in detail in a future paper (Parfrey 2017, in preparation).   

The disc region is defined by its boundary curve, which we take to be a parabola,
\begin{equation}
  Z^2\lp R \rp = \Delta Z \left( R - \rmag\right), \,\, \mathrm{for} \,\, R \geq \rmag,
  \label{eq:parabola}
\end{equation}
where $(R,Z)$ are the cylindrical radius and height, and $\Delta Z$ sets the disc thickness; this profile is chosen solely as a simple one-parameter family of smooth curves, and is not intended to reflect any particular disc model (although see Section~\ref{sec:probSpec}). Inside the disc, the current density is set by requiring that Ohm's law, $\vec{\tilde{J}} = \sigma \vec{\tilde{E}}$, holds in the fluid rest frame. Following Lorentz transforms of the fields from the fluid frame to the static inertial (`lab') frame, $(\vec{\tilde{J}}, \vec{\tilde{E}}, \vec{\tilde{B}}) \rightarrow (\J_{\rm disc}, \vec{E}, \vec{B})$, the current density is given by
\begin{equation}
  \J_{\rm disc} = \Gamma \sigma \ls \E + \vec{v}\times\B/c - \lp \vec{v}\cdot\E\rp \vec{v}/c^2 \rs + \rhoe \vec{v} ,
  \label{eq:Jdisc}
\end{equation}
where $\sigma$ is the conductivity in the fluid rest frame and $\Gamma = 1/\sqrt{1 - v^2}$. The choices of the disc's velocity and conductivity fields, $\vec{v}$ and $\sigma$, will be described in the following sections. The current blends from $\J_{\rm FFE}$ to $\J_{\rm disc}$ in a thin skin at the disc's surface: $\J = Q \J_{\rm FFE} + (1-Q)\J_{\rm disc}$ where $0 < Q < 1$ in the skin.

When $\rmag < \rco$ the pulsar should be accreting, and so there should be another component to the accretion flow: a stream of material leaving the disc plane where the star's magnetic field disrupts the disc, and being funneled along closed field lines to the star's magnetic poles. 
  In this region the accreting plasma's angular momentum is transferred to the star via the `pulling forward' of the star's magnetic field lines. The transitional region at the funnel's base in the disc is neither strongly magnetically dominated nor matter dominated, and so cannot be directly captured in our force-free simulations. We simply omit the accretion funnel from the simulations, on the grounds that the behaviour of the accretion stream should not significantly affect the quantity in which we are most interested, the spin-down torque applied by open field lines, if the disc is truncated well inside the corotation radius. 
 
One can take our disc truncation radius $\rmag$ as the innermost point at which the accretion flow is close to Keplerian; inside $\rmag$ the funnel flow begins and the star's magnetic field brings the gas to corotation. The spin-up torques we calculate therefore include only that part due to field lines entering the Keplerian disc inside $\rco$, and do not include the torque applied by the accreting matter in the funnel flow, $\Nacc \approx \mdot \sqrt{GM\rmag}$. This additional torque can be estimated using equation~(\ref{eq:Nacc}), given a value for the parameter $\xi = \rmag/\ra$.

The pulsar is modeled as an aligned rotator with parallel magnetic moment $\vec{\mu}$ and angular velocity $\vec{\Omega}$, both pointing in the positive cylindrical-$Z$ direction; i.e. $B_{r, *} > 0$  ($<0$) above (below) the equatorial plane and $v_{\phi, *} > 0$, so stellar rotation alone generates $B_{\phi} < 0$ ($>0$) on open field lines rooted in the north (south) poles. The torques $N$ quoted below are calculated by integrating the $r\phi$ component of the Maxwell stress tensor over the stellar surface.

The grid has $N_r \times N_\theta = 512 \times 512$ nodes, over a spherical domain $\rstar \leq r \leq 40\, \rstar$, $0 < \theta < \pi$; a smooth coordinate mapping biases the radial grid resolution towards the star, while the grid is equispaced in $\theta$. To maintain stability, high-order spectral filtering, at eighth and thirty-sixth order, is applied to the electric and magnetic fields at the end of each time step; see \cite{Parfrey:2012ab} for a discussion of the filters (note however that the current sheet capturing procedure described above allows the strength of the eighth-order filtering, and the resulting numerical dissipation, to be decreased by a factor of ten as compared to this earlier paper).


\section{Magnetospheres of pulsars and $\alpha$-discs}
\label{sec:simsAlpha}

Here we describe our most physically complete simulations, in which the disc velocity field and conductivity profile are determined by a basic $\alpha$-disc model. To investigate the effect of the disc's radial accretion velocity, we also performed otherwise identical simulations in which the accretion velocity was set to zero.

\subsection{Problem specification}
\label{sec:probSpec}
To determine the disc properties, we begin with an estimate of the effective viscosity due to tubulence, $\nu = \alpha \cs h$ \citep{Shakura:1973aa}, where $\cs$ and $h$ are the disc's sound speed and half-thickness, and $\alpha$ is expected to be $\sim 0.01$--0.1. Our discs have a parabolic shape, and we set $h(R) = Z(R)$ from equation~(\ref{eq:parabola}). The radial accretion velocity is then given by $v_r = \alpha (h/r)^2 \vk$; the Keplerian azimuthal velocity is $\vk = (\rg/r)^{1/2} c$, and we set $\rg = 0.21 \rstar$, equivalent to $M_* = 1.4 \msun$ and $\rstar = 10$ km. Our primary simulations have disc velocity field $\vec{v} = (v_r, v_\theta, v_\phi) = (v_r, 0, \vk)$, while those without the accretion velocity simply have Keplerian rotation: $\vec{v} = (0,0,\vk)$.

For the disc's conductivity profile, we relate the turbulent viscosity $\nu$ and turbulent magnetic diffusivity $\nu_{\rm m}$ via an effective magnetic Prandtl number, $\Prm = \nu / \nu_{\rm m}$. We take $\Prm = 1$ for all simulations, excepting those discussed in Section~\ref{sec:prandtl}, since both diffusive processes are generated by the turbulence. The conductivity $\sigma$ is found via $\nu_{\rm m} = c^2 / 4\pi\sigma$, giving $4\pi\sigma = \ls \alpha \, (h^2/r)\, \vk \rs^{-1} c^2$, where we have used $\cs = (h/r)\, \vk$. 

\begin{table}
  \centering
  \begin{tabular}{l c c r r}
    Label  &  $\alpha$  &  $\Delta Z \, \left[\rstar\right]$  &  Peak $|v_r|/c$  &  Min.\ $\sigma\, \left[c/\rstar\right]$ \\  
    \hline
    R1  &  0.1  &  0.04  &  $1.8\times 10^{-4}$  &  1751 \\
    R2  &  0.4  &  0.04  &  $7.5\times 10^{-4}$  &  438 \\
    R3  &  0.1  &  0.64  &  $2.9\times 10^{-3}$  &  109 \\
    NoAdv1  &  0.1  &  0.04  &  0  &  1751 \\
    NoAdv2  &  0.4  &  0.04  &  0  &  438 \\
    NoAdv3  &  0.1  &  0.64  &  0  &  109 \\
  \end{tabular}
  \label{tab:reference}
  \caption{ Simulations with $\alpha$-discs: disc parameters and basic properties. $\Delta Z$ sets the disc thickness via equation~(\ref{eq:parabola}). The peak radial accretion velocity occurs at $\sim 2.5\,\rstar$. The conductivity depends weakly on radius beyond this point; the minimum value indicated here occurs at $\sim 4.5 \,\rstar$. }
\end{table}

We performed three reference simulations (R1--3), and three corresponding simulations in which the radial advection contribution to the disc's velocity field was set to zero (NoAdv1--3). The disc parameters, together with the peak accretion velocity and the approximate conductivity over a large range in radius, are given in Table~\ref{tab:reference}. In all cases, the pulsar's light cylinder is set to $\rlc = 16\,\rstar$, corresponding to a spin frequency of about 300~Hz for $\rstar = 10$~km, which places the corotation radius at $\rco \approx 3.75\, \rstar$; the disc's inner edge (the magnetospheric radius) is $\rmag = 1.5 \, \rstar$. This choice of parameters provides good separation between the four radii of interest: $\rstar$, $\rmag$, $\rco$, $\rlc$. 

Our thinner disc profile, with $\Delta Z = 0.04\, \rstar$, approximately reflects an accretion flow having $\mdot = 0.1\, \mdot_{\rm Edd}$, which may be appropriate, for example, for atoll sources in the bright soft state \citep[e.g.][]{Done:2003aa}. At this accretion rate the flow should be radiation-pressure dominated out to roughly $2\, \rlc$, and our parabolic disc shape agrees with the expected profile of such a flow \citep{Shakura:1973aa} to within 25\% inside the light cylinder; the two profiles have equal thickness at about $7\, \rstar$, where $h/r \approx 0.07$. Our thicker profile may be roughly appropriate for the geometrically thicker flows which may exist in the low-luminosity hard states of atolls and accreting millisecond pulsars, and for the thick dense flows of the mildly super-Eddington Z sources.

The initial magnetic field configuration is a potential dipole field centred on the star. Rotation of the star and motion in the disc begin at $t = 0$, and the simulations were run for at least 100 pulsar spin periods, $P = 2\pi/\Omega$.

\subsection{Field opening, relaxation, and stellar torque}
\label{sec:inflation}

\begin{figure*}
  \begin{center}
    \includegraphics[width=\textwidth, trim = 2.0mm 2.0mm 2.0mm 2.0mm, clip]{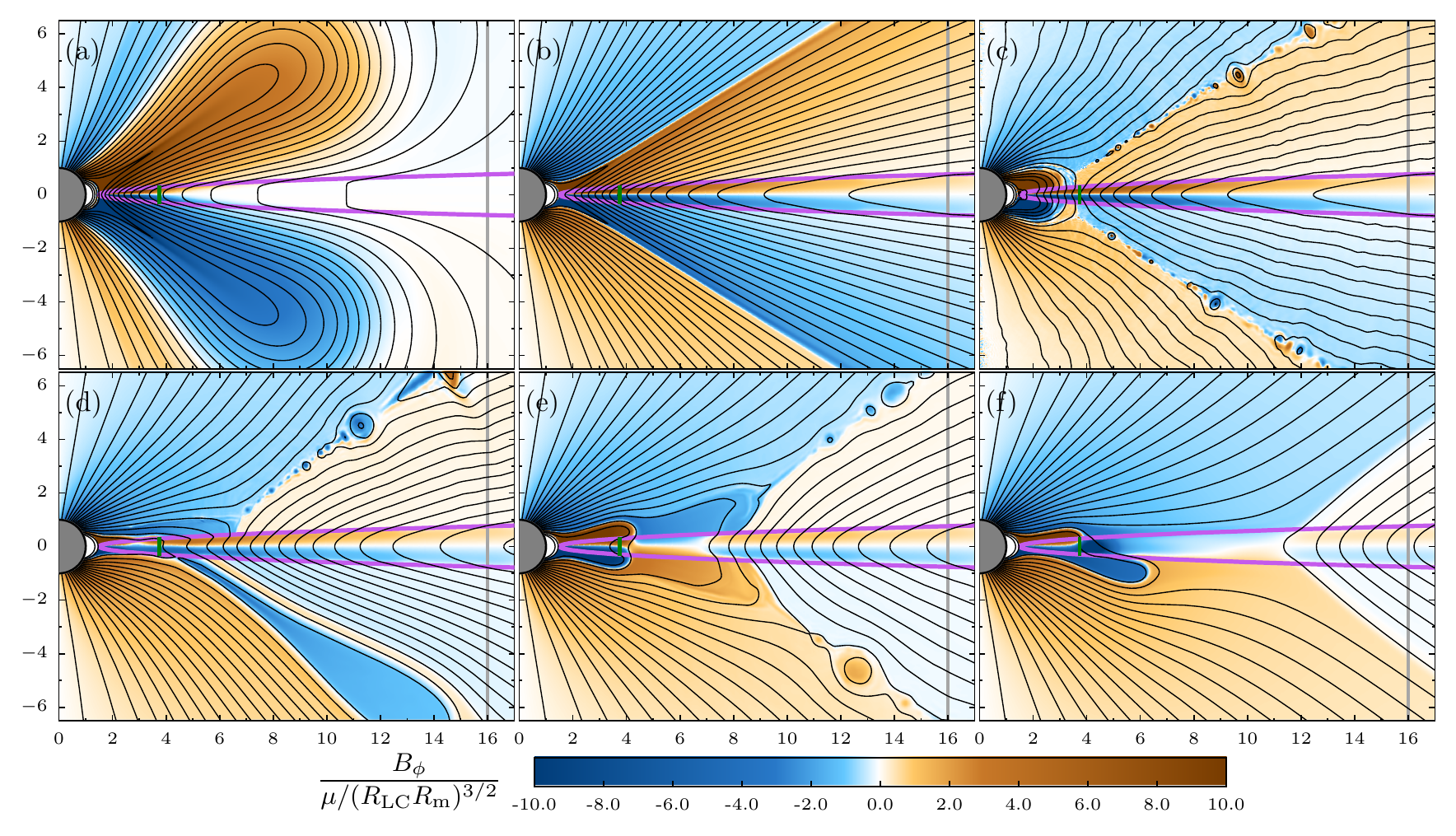}
  \end{center}
  \caption{ \label{fig:multipanelTime}  Evolution of reference simulation R1. The axis scales are in units of $\rstar$. The light cylinder is illustrated with a vertical gray line, while the short green bar indicates the corotation radius in the disc. The outline of the disc is drawn in purple. Twenty poloidal field lines are shown in black, equally spaced in flux function $\ff$ between 0.01 and 0.8. The colour indicates the toroidal magnetic field. (a) $t = 0.4\,P$, (b) $t = 3.2\,P$, (c) $t = 3.4\,P$, (d) $t = 24.4\,P$, (e) $t = 37.8\,P$, (f) $t = 92.2\,P$, where $P$ is the pulsar spin period. }
\end{figure*}

\begin{figure}
  \begin{center}
    \includegraphics[width=3.35in, trim = 2.0mm 2.0mm 2.0mm 2.0mm, clip]{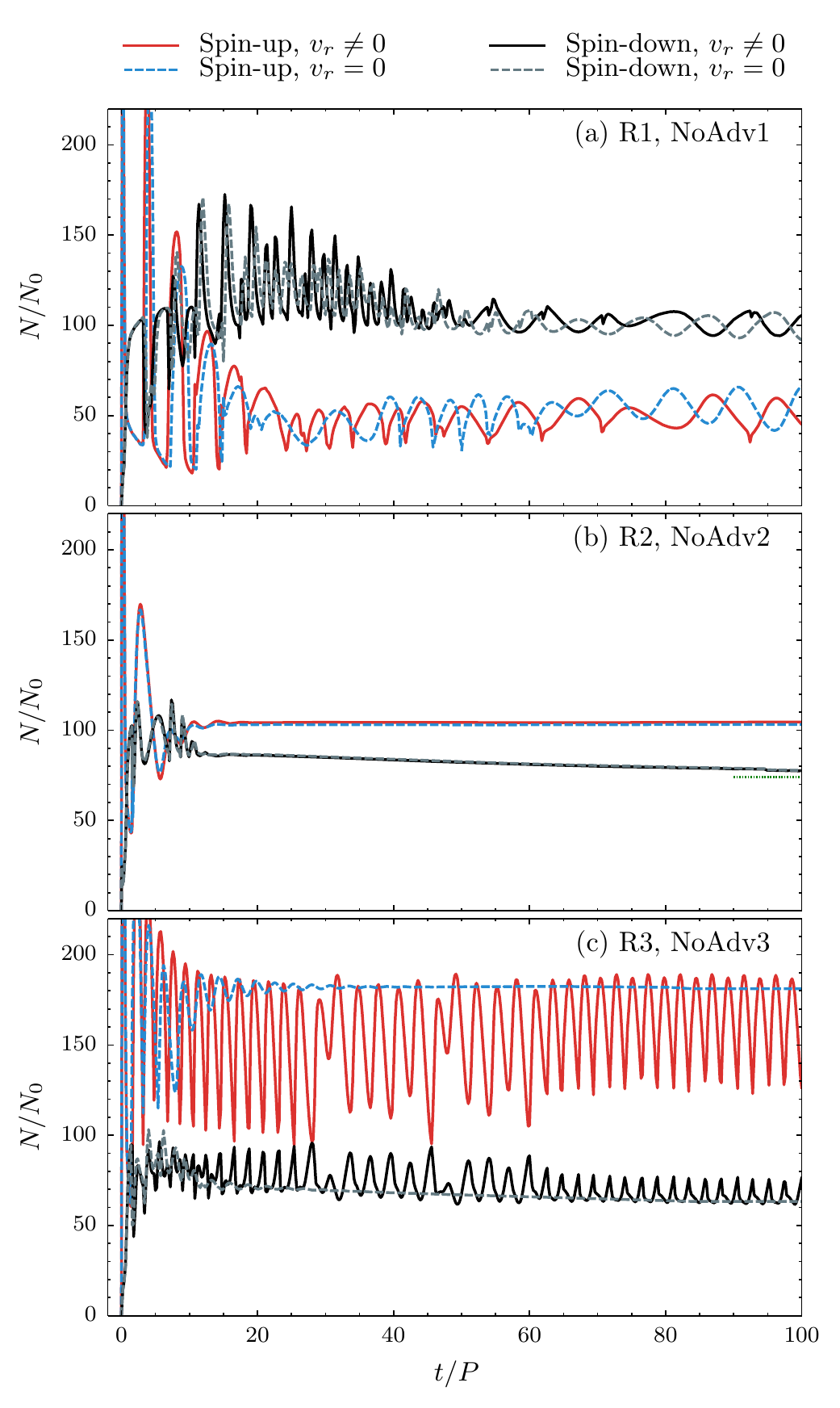}
  \end{center}
  \caption{ \label{fig:torqueOverTime}  Spin-up (red and blue lines) and spin-down (black and grey lines) contributions to the torque $N$ on the star, in units of the spin-down torque on an isolated pulsar $\No$, for the three sets of $\alpha$-disc simulations. Solid lines are drawn for the reference simulations, dashed lines for the `NoAdv' runs in which the disc's accretion velocity was set to zero. The green dotted line in (b) indicates the level at which the spin-down torque stabilizes once the current sheets leave the computational domain, which occurs at $t \approx 120\,P$. }
\end{figure}

The evolution of the R1 ($\alpha = 0.1$, $\Delta Z = 0.04\,\rstar$) simulation is illustrated in Fig.~\ref{fig:multipanelTime}. The corresponding spin-up and spin-down contributions to the torque on the star are plotted (with red and black lines respectively) in Fig.~\ref{fig:torqueOverTime}(a). 

Note that the spin-up torques shown in Fig.~\ref{fig:torqueOverTime} comprise exclusively the torques applied by field lines which enter the Keplerian disc inside the corotation radius; see Section~\ref{sec:method}. There is an additional spin-up torque applied by the plasma accreting in a funnel from the magnetospheric radius, which can be estimated with equation~(\ref{eq:Nacc}). Using the $\rmag$ and $\rlc$ values from the simulations, one finds that $\Nacc/N_0 \approx 858\, \xi^{7/2}$, or $\Nacc/N_0 \approx$ 76 for $\xi =$ 0.5. 

When the simulation begins, differential rotation between the star and the disc twists the field lines, generating toroidal magnetic field whose pressure pushes the field lines outwards. The most rapidly twisted field lines are those entering the inner disc, inside $\rco$, so these power this first phase of poloidal inflation, Fig.~\ref{fig:multipanelTime}(a); toroidal field of the opposite sign to that in the expanding bubble is seen on those field lines frozen into the star nearer the poles, which is due to sweep-back of the field lines entirely by the stellar rotation. The forward-twist at the star of the field lines in the inflating bubble deliver a powerful spin-up torque. 

As the apices of the expanding field lines move outwards at nearly the speed of light, their footpoints on the star and disc can no longer communicate via Alfv\'{e}n waves. There are effectively two sets of open field lines, those frozen into the star and those entering the disc, with each set supporting a single sign of $B_\phi$, corresponding to the rotational sweep-back; see Fig.~\ref{fig:multipanelTime}(b). The jump of $B_r$ and $B_\phi$ between the two sets of open field lines is supported by a strong current layer, which becomes thinner (i.e. has increasing current density) as the field lines inflate. There is also a small amount of closed flux which continues to couple the inner disc to the star, but here its spin-up contribution to the torque is only roughly a third of the spin-down torque applied by the much larger bundle of open field lines.

Eventually the current sheet becomes unstable to tearing-mode reconnection, and some of the open flux reconnects, producing a hierarchy of plasmoids which are ejected from the system, Fig.~\ref{fig:multipanelTime}(c). Here the first reconnection event begins at $t \approx 3.3\,P$. The newly closed field lines, once again joining the star to the disc, are immediately subject to rapid twisting and begin to re-inflate, leading to a quasi-periodic cycle of flux inflation and reconnection, Fig.~\ref{fig:multipanelTime}(d). The stellar torque oscillates sharply over one of these cycles, with the spin-down torque being largest, and the spin-up contribution smallest, just before reconnection.

At all times, the disc's open poloidal field lines enter in a pushed-forward, `$<$'-like configuration due to magnetic pressure balance. Sequences of quasi-static twisted force-free equilibria expand at an angle of $\sim 60^\circ$ to the vertical \citep{Lynden-Bell:1994aa}. When the toroidal field on the disc's open flux is much less than that on the stellar flux, one can think of the disc field as being pushed forward by the pulsar wind, which would otherwise become nearly radial. The consequence is that, when radial field of opposite signs annihilates across the disc due to its finite conductivity, there is a slow drift outwards of the points at which the field lines enter the disc; see Section~\ref{sec:expulsion} for details. In this simulation the resistive drift velocity is larger than the radial accretion velocity, and the disc's flux is gradually expelled from the system, Fig.~\ref{fig:multipanelTime}(c)--(f). 

A small amount of flux remains trapped in the disc near the corotation radius, where the shear between star and disc is small. However the disc has opened most of the field lines which initially entered it inside $\rco$, and expelled their disc footpoints to large radii; the trapped flux originally entered near the disc's inner edge.  By $t = 100\,P$ most of the disc's open flux has been expelled through the light cylinder, and the inner magnetosphere has settled into a quasi-steady state. 

The opening and reconnection cycles of the remaining star--disc coupling field continue, causing the torques to oscillate, though with the spin-down torque being always greater than the spin-up torque. Once the additional accretion torque is added via equation~(\ref{eq:Nacc}), one can see that this system would be in approximate long-term spin equilibrium if $\xi \sim 0.5$, even though the disc is truncated well inside the corotation radius.
There is no evidence that the system is tending towards a true steady state, i.e. one in which $\partial_t \vec{B} = 0$ and the torques are constant in time.

\begin{figure}
  \begin{center}
    \includegraphics[width=3.35in, trim = 2.0mm 2.0mm 2.0mm 2.0mm, clip]{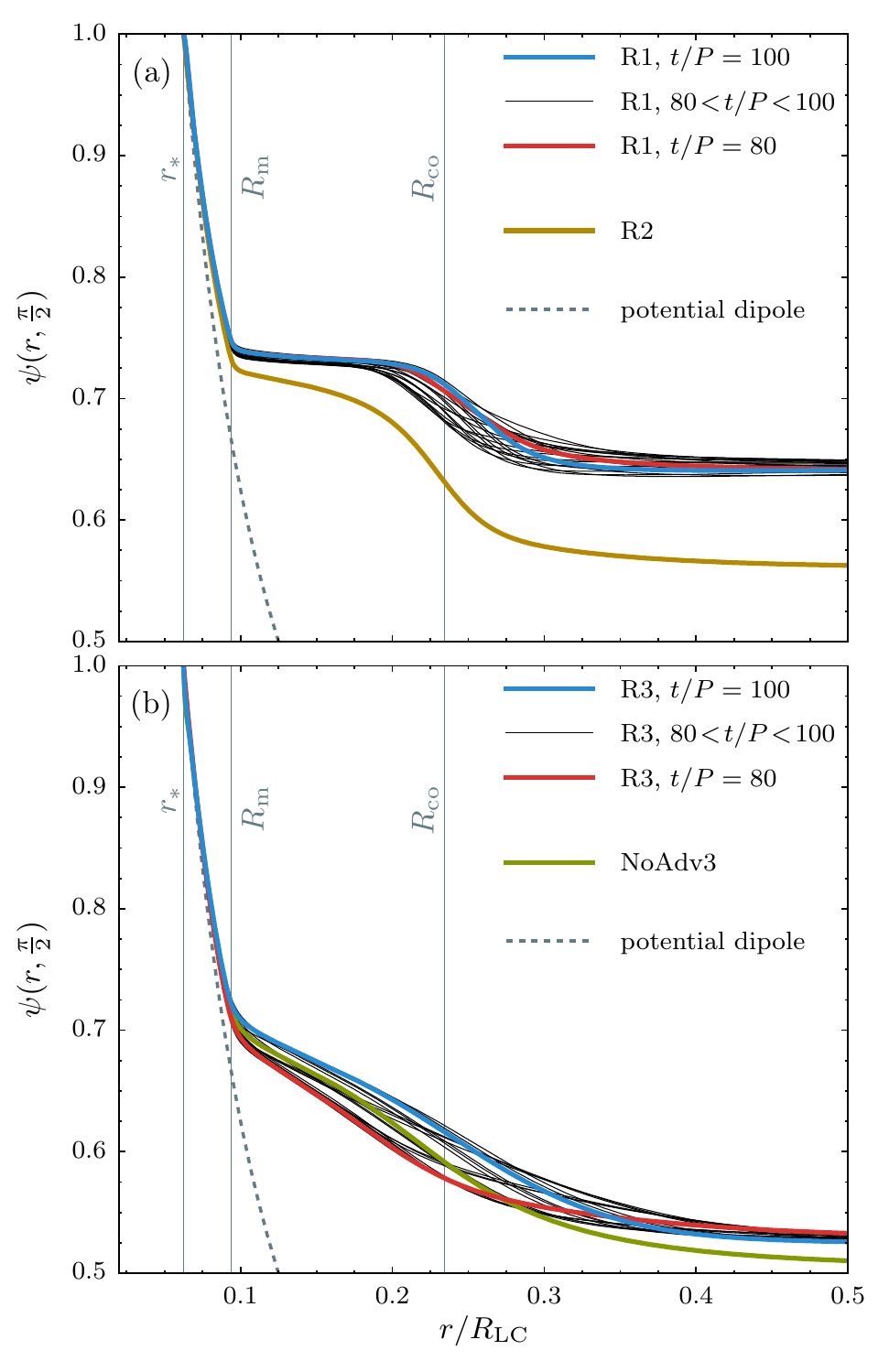}
  \end{center}
  \caption{ \label{fig:ffAlpha}  Normalized flux function [$\ff(\rstar,\pi/2) = 1$] in the equatorial plane. The dashed grey line denotes the flux function for the initial dipole field. The disc extends to the right of the $\rmag$ line. (a) The thick red and blue lines are drawn for the R1 ($\alpha = 0.1$) simulation at $t/P = 80$ and 100 respectively; thin black lines indicate the solution at twenty equally spaced times in between. The R2 ($\alpha = 0.4$) simulation is represented in its final steady state by the yellow line. (b) As above, but for the thick-disc R3 simulation, and the equivalent run without accretion velocity, NoAdv3, whose steady state is illustrated in green. }
\end{figure}

The behaviour when the disc's radial accretion velocity $v_r$ is set to zero (simulation NoAdv1) is similar, indicating that for this $\alpha$-disc model accretion is too slow to have much effect. The disc's open flux is expelled outwards slightly faster, and some of the torque oscillations are not as pronounced [see the dashed blue and grey lines in Fig.~\ref{fig:torqueOverTime}(a)].

The distribution of the poloidal flux function $\ff$ through the inner part of the disc at late times is shown in Fig.~\ref{fig:ffAlpha}(a). The poloidal magnetic field is proportional to gradients of $\ff$, $\vec{B_{\rm P}} = \nabla\ff\times\vec{\hat{\phi}}/2\pi r\sin\theta$, so the plateaus indicate regions of weak poloidal magnetic field in the disc. The $\ff$ distribution oscillates within a restricted range over an opening and reconnection cycle. Most of the poloidal flux is concentrated near the corotation radius; depending on the phase of the reconnection cycle, there can be more flux entering the disc inside or outside corotation (the flux through the equator between any two radii is given by the difference in $\ff$ between them).

\subsection{Flux expulsion versus inward dragging by accretion}
\label{sec:expulsion}

We can estimate the outward resistive drift velocity of the field lines, following \cite{van-Ballegooijen:1989aa}. Working in cylindrical coordinates, the non-ideal terms in the induction equation for the vertical magnetic field are given by $\partial_t B_Z = \cdots + (1/R) \partial_R \ls R \nu_{\rm m} \lp\partial_R B_Z - \partial_Z B_R \rp \rs$, where $\nu_{\rm m}$ is the magnetic diffusivity; the first term corresponds to radial diffusion of $B_Z$, the second to change of $B_Z$ due to vertical annihilation of $B_R$. The annihilation term should dominate as long as the disc height $h$ is the smallest length scale available, in which case the radial drift speed is $v_{\rm resist} \sim  (\nu_{\rm m}/h) |B_R/B_Z| = (\nu_{\rm m}/h) \tan\theta_{\rm B}$, where $\theta_{\rm B}$ is the angle to the vertical at which the field enters the disc. 

Equivalently, one can think of the field lines as diffusing outwards with the speed of diffusion across the disc, $\nu_{\rm m}/h$, times the geometric factor which converts this vertical diffusion into apparent radial movement, $|B_R/B_Z|$.

The effective turbulent magnetic diffusivity can be expressed as $\nu_{\rm m} = \nu / \Prm = \alpha \cs h / \Prm$, and so $v_{\rm resist} \sim \alpha \cs \tan\theta_{\rm B} / \Prm$. The radial accretion speed is given by $v_{\rm accrete} = v_r \sim \alpha (h/r) \cs$, so their ratio is
\begin{equation}
  \frac{v_{\rm accrete}}{v_{\rm resist}} \sim \frac{h}{r} \frac{\Prm}{\tan\theta_{\rm B}}.
  \label{eq:expulsionSpeed}
\end{equation}
Since in the simulations the open field lines enter the disc with $\tan\theta_{\rm B} \sim 1$ and we expect $\Prm \sim 1$, the competing radial speeds are simply related by $v_{\rm accrete}/v_{\rm resist} \sim h/r$. Outward expulsion of the field lines should easily overcome the radial accretion flow for thin discs, as is seen in the simulation described above, and the two should be of roughly equal strength for thicker discs, where the $\alpha$-disc model in any case ceases to be a reliable approximation.

\subsection{Varying the $\alpha$ parameter: more diffusive discs}
The specification of the R2 simulation is identical to the R1 run described above, except with a Shakura-Sunyaev $\alpha$ parameter four times larger, which reduces the disc's conductivity, and increases its radial accretion velocity, by a factor of four (see Table~\ref{tab:reference}). The spin-up/spin-down torques on the star are shown in Fig.~\ref{fig:torqueOverTime}(b). The early phases are very similar to the R1 simulation, but after several phases of opening and reconnection the inner magnetosphere reaches a steady state and the spin-up torque, caused by field lines entering the disc inside $\rco$, becomes constant. 

The open field lines rooted in the disc are slowly expelled outwards, and the spin-down torque on the star declines as the stellar open field lines relax toward their final nearly radial configuration. The poloidal deflection of the stellar field lines by the disc's magnetic field only slightly increases the power extracted in the pulsar wind. The innermost open field line attached to the disc moves through the light cylinder at $t\approx 37\,P$, and exits the computational domain at $t\approx 120\,P$, at which time spin-down torque also becomes constant and the entire magnetosphere settles into a steady state. The final spin-down torque is about $74\, \No$, where $\No$ is the torque on an isolated pulsar of the same rotation rate; it is roughly 17\% larger when the inner magnetosphere becomes approximately steady, at $t\approx 15\,P$.

In this simulation, the final relative strengths of the spin-up and spin-down contributions to the stellar torque have been reversed when compared to R1. The lower disc conductivity has allowed more field lines to remain closed in the disc, especially inside $\rco$, which has increased the spin-up torque and decreased the spin-down torque. The radial accretion velocity is unimportant in this simulation, as can be seen in Fig.~\ref{fig:torqueOverTime}(b), where the dashed lines of the $v_r = 0$ run (NoAdv2) lie very close to the solid lines of R2 throughout. 

The flux function distribution in the disc for the R2 run's final steady state is shown in Fig.~\ref{fig:ffAlpha}(a). As one would expect, more flux threads the disc when its diffusivity is increased. This is true particularly inside corotation, while the flux threading the outer disc does not change as dramatically. The region around $\rco$ in which the poloidal flux is concentrated has broadened in comparison to the lower diffusivity R1 simulation.

\subsection{Thicker discs}

\begin{figure}
  \begin{center}
    \includegraphics[width=3.35in, trim = 2.0mm 2.0mm 1.0mm 2.0mm, clip]{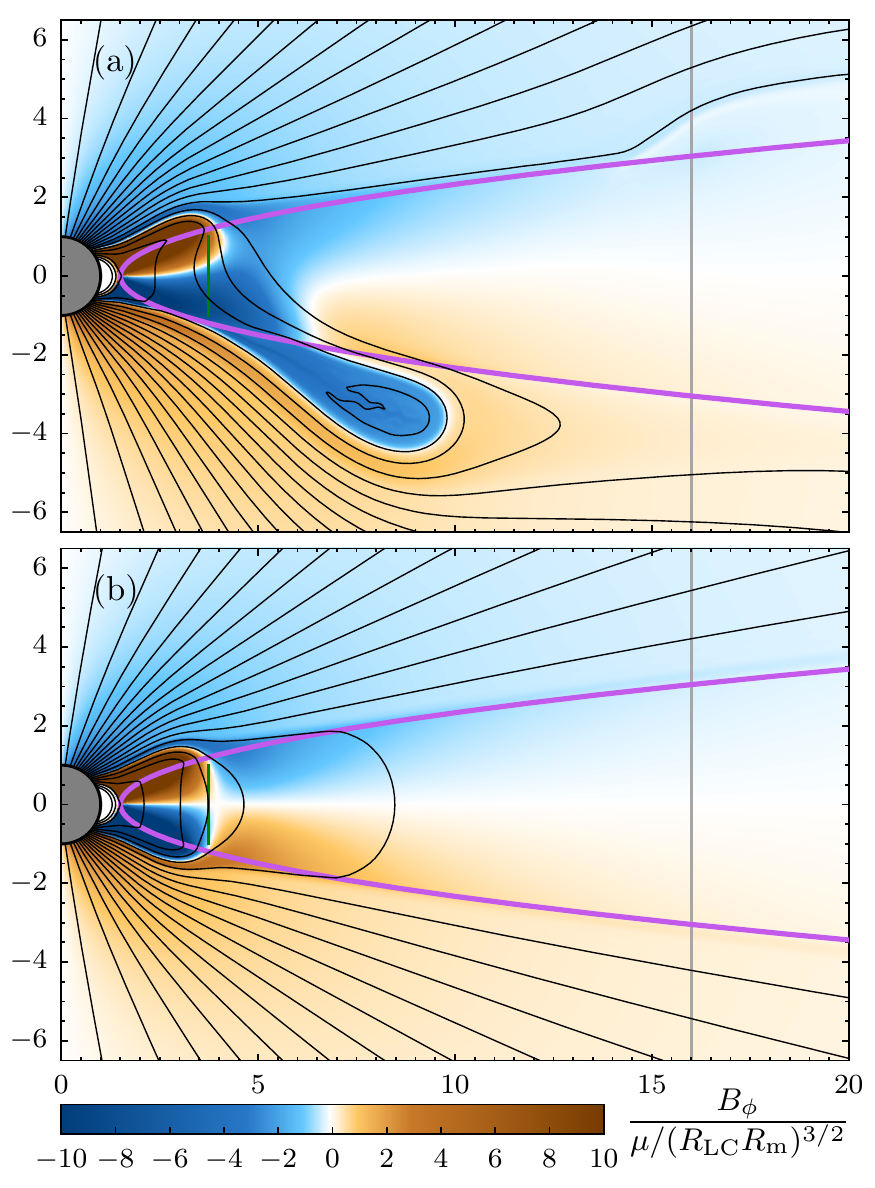}
  \end{center}
  \caption{ \label{fig:thickDisc} Thick-disc simulations at $t = 100\,P$. (a) Full disc model, run R3: opening and reconnection from the inner disc continue indefinitely, giving torque oscillations; (b) no radial accretion velocity, run NoAdv3: a steady state is reached and the torques are constant. Markings are as in Fig.~\ref{fig:multipanelTime}.}
\end{figure}

The final $\alpha$-disc pair, R3 and NoAdv3, were performed to investigate the behaviour of pulsar magnetospheres coupled to hotter, partially pressure-supported, discs. Here the assumptions underlying the standard $\alpha$-disc model do not apply, but it usefully allows a straightforward comparison with the thin-disc simulations, and reflects the more important features of thick discs for our purposes: the accretion velocity increases and the conductivity decreases as the disc becomes thicker. The R3 simulation has $\alpha = 0.1$, as in R1, but the disc is four times thicker, and so $v_r$ and $\sigma$ are, respectively, sixteen times larger and smaller.

The reference run, R3, behaves in many respects like our fiducial simulation R1: the disc's open field lines are gradually driven outward, eventually well beyond the light cylinder, but opening and reconnection of those field lines entering the inner disc continues indefinitely. The stellar torques undergo large oscillations due to the opening and reconnection cycle, whose maintenance requires radial accretion in the disc; see Fig.~\ref{fig:torqueOverTime}(c). When the radial velocity is removed in run NoAdv3, the inner disc settles into equilibrium, the spin-up torque becomes nearly constant, and the spin-down torque decreases slowly as in run R2. These two outcomes are illustrated in Fig.~\ref{fig:thickDisc}. 

Without the accretion velocity, the spin-up torque remains close to the highest points of the full model's spin-up oscillations, and the spin-down torque stays near the lowest points of the corresponding variations. This is because the oscillations are primarily caused by the opening of field lines coupling to the disc inside corotation, as can be seen in Fig.~\ref{fig:thickDisc}(a). Therefore the opening of this flux, due to the presence of the disc radial velocity in the full model, decreases the spin-up torque in comparison with the model without disc radial advection (fewer field lines dragging the star forwards) and similarly increases the spin-down torque (stronger pulsar wind from larger stellar open flux). The thick disc's lower conductivity allows more field lines to remain closed inside $\rco$, diffusing azimuthally through the disc material, and the spin-up torque is higher than in the simulations with thinner discs. 

The poloidal flux through these thick discs is shown in Fig.~\ref{fig:ffAlpha}(b). The inner plateau of $\ff$ is now much less evident, with the stellar magnetic field being distributed more evenly through the disc due to its reduced conductivity (see Table~1). The opening and reconnection cycle is visible in the clustering of the R3 simulation's lines into several distinct bands. The total open flux is less than in the simulations with thinner, less diffusive, discs, but not dramatically so. When the accretion velocity is removed (NoAdv3), in the inner disc the equatorial flux distribution lies within the range of excursion of the full simulation over its opening cycle. Further out, the opening events from the inner disc in the R3 simulation allow more field lines to remain open in the quasi-steady state, since their relaxation time is longer than the opening cycle; this accounts for the reduced flux through the disc in the R3 simulation, compared to NoAdv3, during all phases of the opening cycle for $r/\rlc \gtrsim 0.3$.

\subsection{Effect of increased radial accretion velocity}
\label{sec:prandtl}

\begin{figure}
  \begin{center}
    \includegraphics[width=3.35in, trim = 2.0mm 2.0mm 2.0mm 2.0mm, clip]{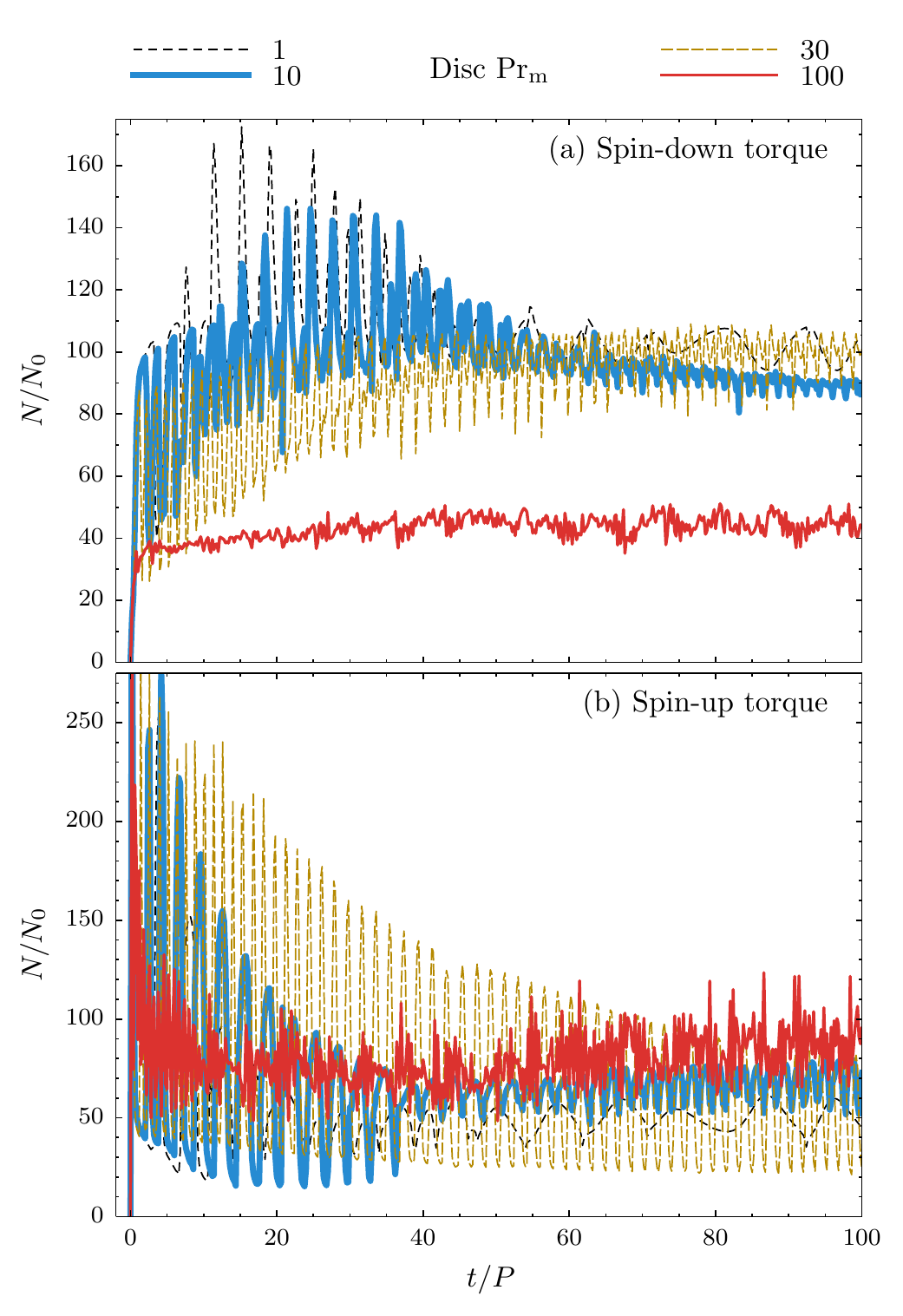}
  \end{center}
  \caption{ \label{fig:prandtl}  (a) Spin-down and (b) spin-up contributions to the torque $N$ on the star, in units of the spin-down torque on an isolated pulsar $\No$, for discs with increased accretion velocity. The disc in each case has a radial velocity which is a factor of $\Prm$ larger than that in the R1 simulation, but is otherwise identical. }
\end{figure}

\begin{figure*}
  \begin{center}
    \includegraphics[width=\textwidth, trim = 2.0mm 2.0mm 2.0mm 2.0mm, clip]{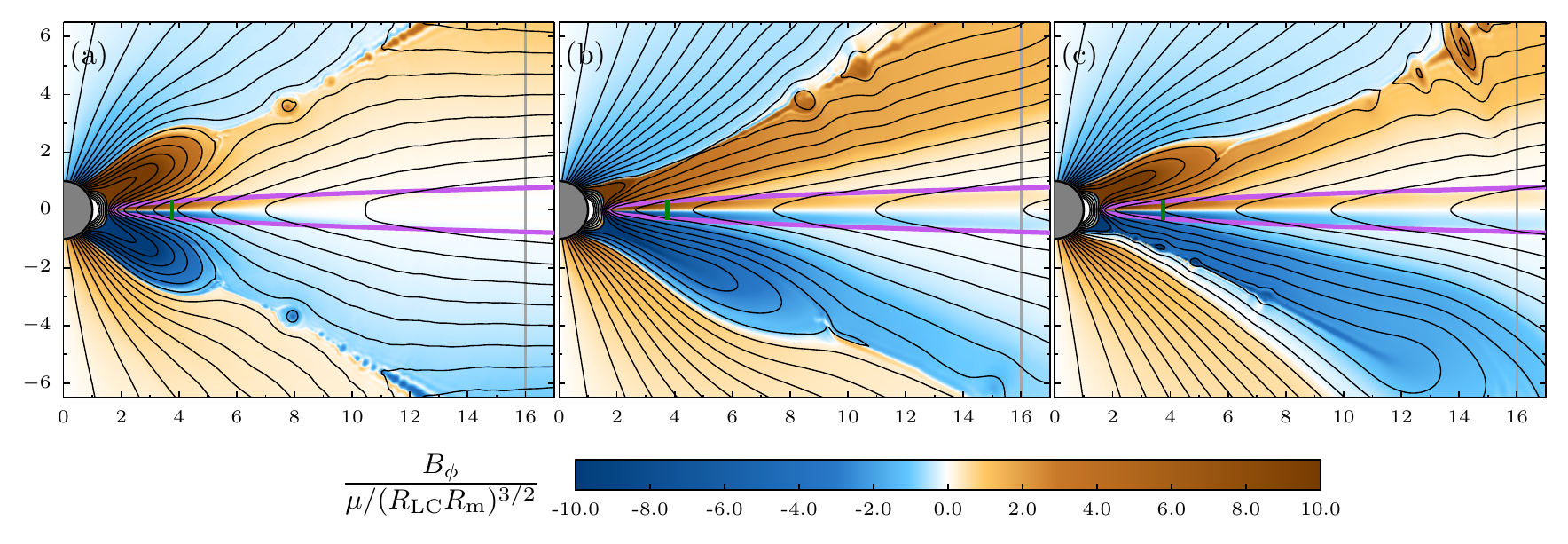}
  \end{center}
  \caption{ \label{fig:multipanelPrm100}  Simulation with strongly increased accretion velocity: $\Prm = 100$. Poloidal field lines and all markings are identical to those of Fig.~\ref{fig:multipanelTime}. (a) $t = 0.6\,P$, (b) $t = 50\,P$, (c) $t = 100\,P$, where $P$ is the pulsar spin period. }
\end{figure*}

In the simulations described above, we set the disc's magnetic diffusivity equal to its viscosity in the $\alpha$-disc model which provides the conductivity and radial accretion velocity to be used in constructing the disc-zone current, equation~(\ref{eq:Jdisc}). Setting the effective turbulent magnetic Prandtl number, $\Prm = \nu/\nu_{\rm m}$, to unity in this manner should generally be a good approximation for MRI-turbulent accretion flows. However, in the case of accretion onto a magnetized star there is an additional component: when stellar field lines are opened due to interaction with the disc, a magneto-centrifugal wind can be driven from the disc, via the part of the (now open) field lines which are rooted there \citep[e.g.][]{Lovelace:1995aa}. This wind extracts angular momentum from the disc, leading to an increased accretion velocity, which can be thought of as an increase of the effective total magnetic Prandtl number above unity. 

We have run a set of simulations with increased $\Prm =$ 10, 30, and 100 to study the influence of this effect on the magnetospheric structure and stellar torques. The disc in each case is identical to that in our fiducial run R1 (see Table~1), except that the radial accretion velocity is increased by a factor of $\Prm$.

The spin-down and spin-up torques on the star in each case are shown in Fig.~\ref{fig:prandtl}; the $\Prm = 1$ curves are identical to those for the R1 simulation in Fig.~2(a). The $\Prm = 10$ simulation behaves generally similiarly to the unity-$\Prm$ case, although the torque oscillations are on a much shorter timescale at late times. For $\Prm = 30$, the outward resistive drift velocity is only slightly larger than the radial accretion velocity over much of the disc (Section~\ref{sec:expulsion}), leading to large torque oscillations over the opening and reconnection cycle; here the field lines diffuse outward when they are open, only to be dragged back nearly to where they began when they are closed following reconnection. The average torque values are roughly similar to those in the $\Prm = 1$ case. 

A different behaviour is seen when the magnetic Prandtl number is increased further, to 100. The magnetosphere is shown at three times in Fig.~\ref{fig:multipanelPrm100}. Now the radial accretion velocity in the inner disc is large enough to sweep inward some of the field lines which initially threaded it, increasing the number of closed, untwisted field lines inside the magnetospheric radius. Because this leads to fewer field lines being opened, the spin-down torque is lower than in the other simulations. Field lines very slowly diffuse outward in the outer disc over the course of the simulation, while those in the inner disc are effectively pinned by the large accretion velocity. The increase in flux threading the inner disc gives a larger spin-up torque than in the other cases. The opening and reconnection cycles are very short, less than a stellar spin period --- the amount of open flux therefore remains nearly constant, and the field lines stay approximately radial at all times, both of which lead to a nearly constant spin-down torque. Although the magnitude of this torque is smaller than in the low-$\Prm$ cases, it is still significantly larger than the isolated-pulsar value.


\section{Open flux and spin-down torque: dependence on magnetospheric radius and disc conductivity}
\label{sec:simsGrid}

To investigate systematically the relationship between open flux and spin-down torque, we performed a grid of thirty-two simulations having the same disc profile but where the magnetospheric radius was placed at eight locations from $\rmag = 1.5$--14 $\rstar$, and the disc conductivity was set in each case to one of four constant values, $4\pi\sigma =$ [400, 800, 1200, 1600] $c/\rstar$.  The light cylinder is at $\rlc = 16\,\rstar$ as in Section~\ref{sec:simsAlpha}, and so $\rco \approx 3.75\,\rstar$.

\begin{figure}
  \begin{center}
    \includegraphics[width=3.35in, trim = 1.0mm 0.0mm 1.0mm 1.0mm, clip]{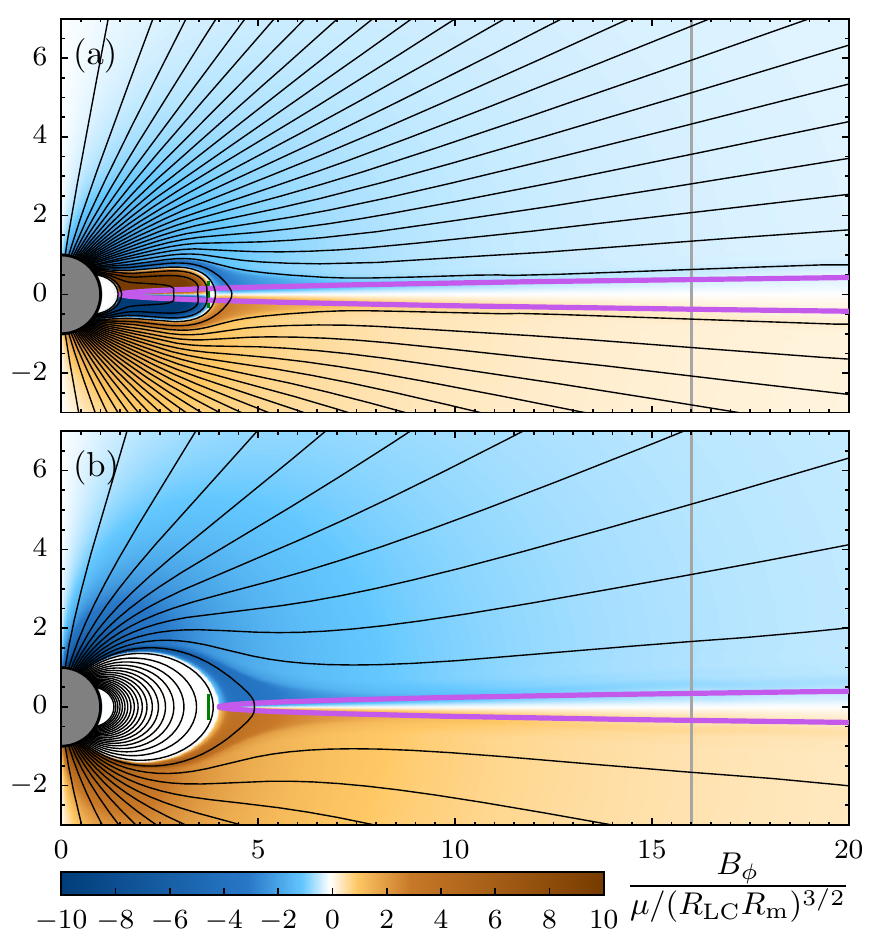}
  \end{center}
  \caption{ \label{fig:steadyState} Steady-state solutions for thin constant-conductivity discs, $4\pi\sigma = 1600\,c/\rstar$. (a) $\rmag = 1.5\,\rstar$, an `accreting' system as $\rmag < \rco$; (b) $\rmag = 4 \,\rstar > \rco$, non-accreting system. Thirty poloidal field lines are equispaced in flux function between $\ff =$ 0.1--0.75; markings are otherwise as in Fig.~\ref{fig:multipanelTime}.}
\end{figure}

\begin{figure}
  \begin{center}
    \includegraphics[width=3.35in, trim = 2.0mm 2.0mm 2.0mm 2.0mm, clip]{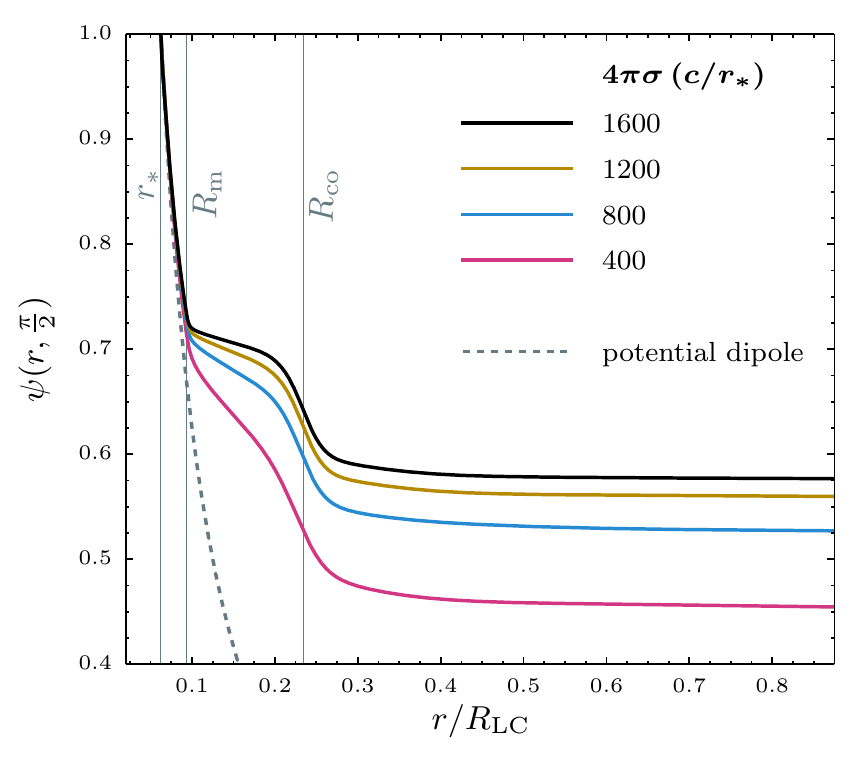}
  \end{center}
  \caption{ \label{fig:gridDiscFlux} Poloidal flux function in the equatorial plane for the four steady states having $\rmag = 1.5\,\rstar$; the black curve (top) corresponds to the same simulation as Fig.~\ref{fig:steadyState}(a). }
\end{figure}

We chose the same thin disc profile for every simulation, having $\Delta Z = 0.01\,\rstar$, which is half as thick as the disc in Section~\ref{sec:simsAlpha}'s R1 and R2. The thinner profile increases the rate of magnetic annihilation across the disc, and so all the simulations settle eventually into stable equilibria in the manner of simulation R2 above, even at the highest conductivity; this is beneficial as the steady states  yield clean measurements for all the quantities of interest. For this disc thickness, the conductivities used roughly correspond to $\alpha$ parameters of $\alpha =$ [1.8, 0.9, 0.6, 0.45]. The radial accretion velocity is comparatively unimportant for these thin discs, and so we have set it to zero, leaving only Keplerian rotation in the imposed velocity field. We also ran an isolated-puslar simulation with the same stellar rotation rate, from whose final steady state we extracted the poloidal flux function at the equator, $\ff_0(r, \pi/2)$, and the total amount of open flux, $\ffopeno$.

Two of the final steady states, with the disc well inside and just outside the corotation radius, are shown in Fig.~\ref{fig:steadyState}, and the distribution of poloidal flux in the disc as a function of the conductivity is illustrated in Fig.~\ref{fig:gridDiscFlux}. There is more poloidal flux closed inside the corotation radius as the disc conductivity is decreased, but in all cases there is very little closed flux much beyond $\rco$.

\begin{figure}
  \begin{center}
    \includegraphics[width=3.35in, trim = 2.0mm 1.0mm 1.0mm 2.0mm, clip]{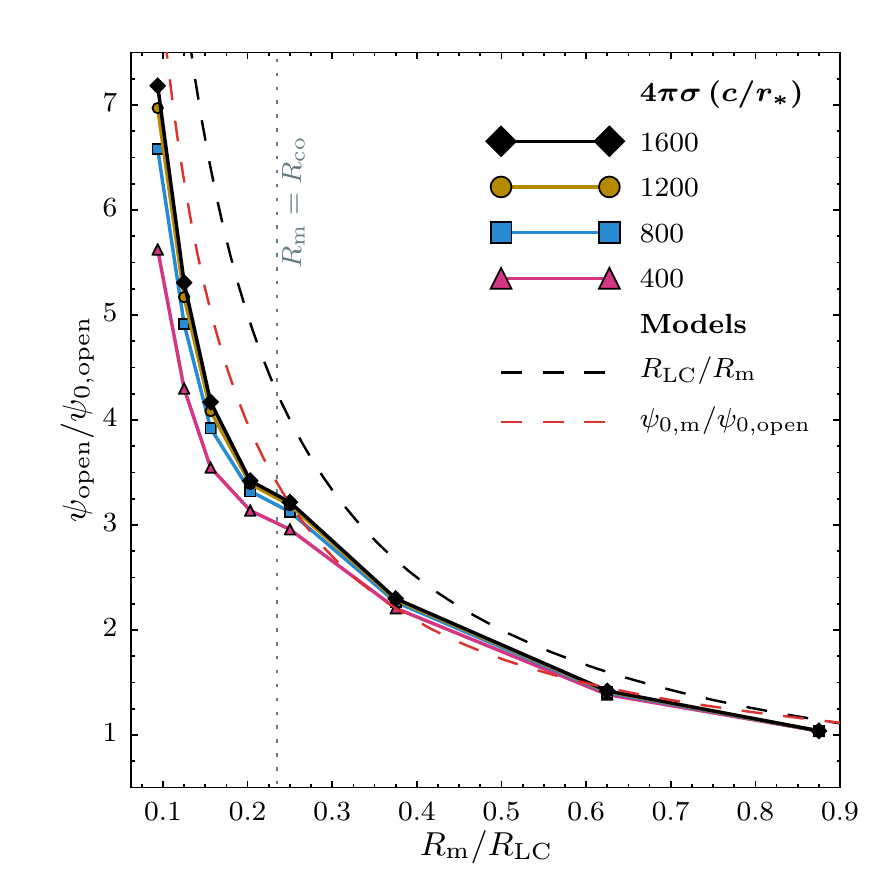}
  \end{center}
  \caption{ \label{fig:flux} Open flux $\ffopen$ in units of its value for an equivalent isolated pulsar $\ffopeno$, for discs of a range of magnetospheric (truncation) radii $\rmag$, for four values of the disc conductivity $\sigma$. The dashed curves show two models: $\ffopen/\ffopeno = \rlc/\rmag$ (black) and $\ffopen/\ffopeno = \ff_{\rm 0, m}/\ff_{\rm 0, open}$ (red), where $\ff_{\rm 0, m}$ and $\ff_{\rm 0, open}$ are the flux function values at the magnetospheric radius and light cylinder, respectively, for the equivalent isolated pulsar. The vertical dotted line indicates the point at which the disc terminates at the corotation radius. }
\end{figure}

The open flux in the final equilibrium state, $\ffopen$, is plotted against the inner disc truncation radius in Fig.~\ref{fig:flux}, for each disc conductivity level. Also shown are two models: in black, the most basic estimate
\begin{equation}
  \ffopen = \frac{\rlc}{\rmag} \ffopeno
\end{equation}
(i.e. equation~\ref{eq:openFluxModel} with $\zeta = 1$), and in red the improved version
\begin{equation}
  \ffopen = \ff_{\rm 0, m},
  \label{eq:ffopenSim}
\end{equation}
where the poloidal flux distribution inside the closed zone has been taken directly from the isolated-pulsar simulation rather than approximated as a potential dipole; as before, it is assumed that all of the flux beyond $\rmag$ is opened ($\zeta = 1$). The open flux has converged, or is close to convergence, as the disc conductivity is increased for all but the innermost truncation radii. The curve kinks slightly when $\rmag$ is just inside $\rco$, where the small amount of strongly twisted flux inside $\rco$ has difficulty opening flux due to the enveloping weakly sheared field lines; this effect becomes less important as $\rmag$ decreases and the amount of flux entering the disc inside $\rco$ increases.

The simplest model, $\ffopen \propto \rlc/\rmag$, is a decent approximation to the open flux for $\rmag \gtrsim \rlc/3$, and equation~(\ref{eq:ffopenSim}) remains good for all magnetospheric radii, especially considering that the closest $\rmag$ to the stellar surface hasn't fully converged in $\sigma$. In other words, the disc can efficiently open field lines even when it is truncated significantly inside the corotation radius.

\begin{figure}
  \begin{center}
    \includegraphics[width=3.35in, trim = 1.0mm 1.0mm 1.0mm 2.0mm, clip]{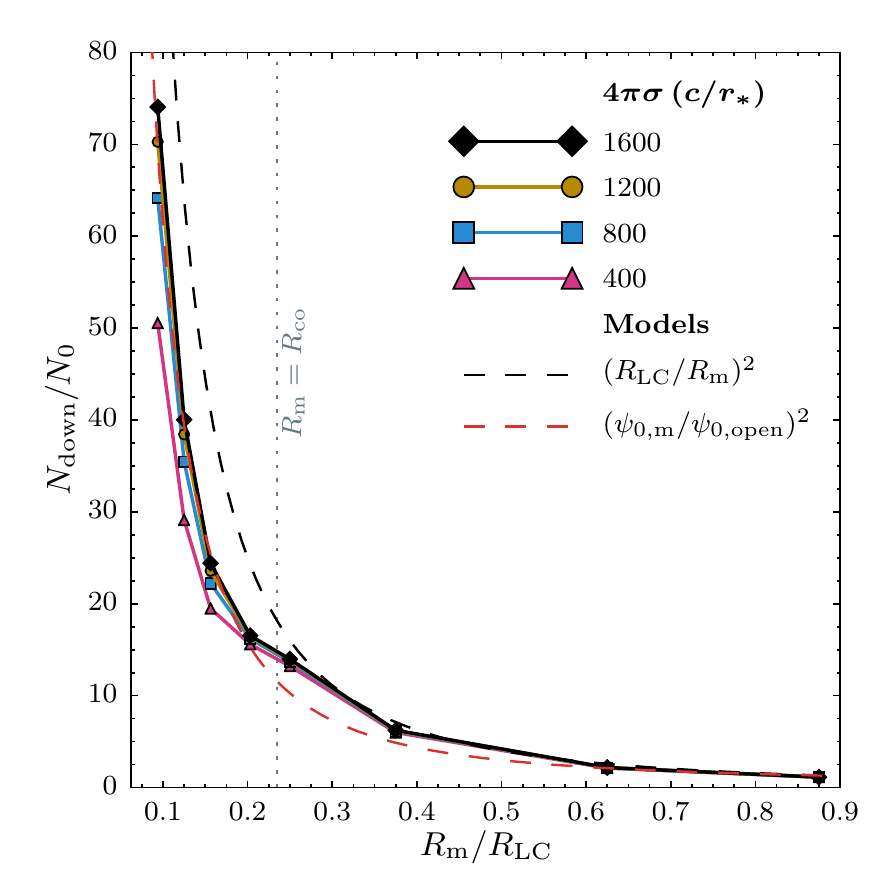}
  \end{center}
  \caption{ \label{fig:torque} Total spin-down torque on the pulsar $\Ndown$ in units of its equivalent isolated pulsar value, $\No = - \mu^2 \Omega^3/c^3$, and the models $\Ndown/\No = (\rlc/\rmag)^2$ (black dashed line) and $\Ndown/\No = (\ff_{\rm 0, m}/\ff_{\rm 0, open})^2$ (red dashed line). }
\end{figure}

\begin{figure}
  \begin{center}
    \includegraphics[width=3.35in, trim = 1.0mm 1.0mm 1.0mm 2.0mm, clip]{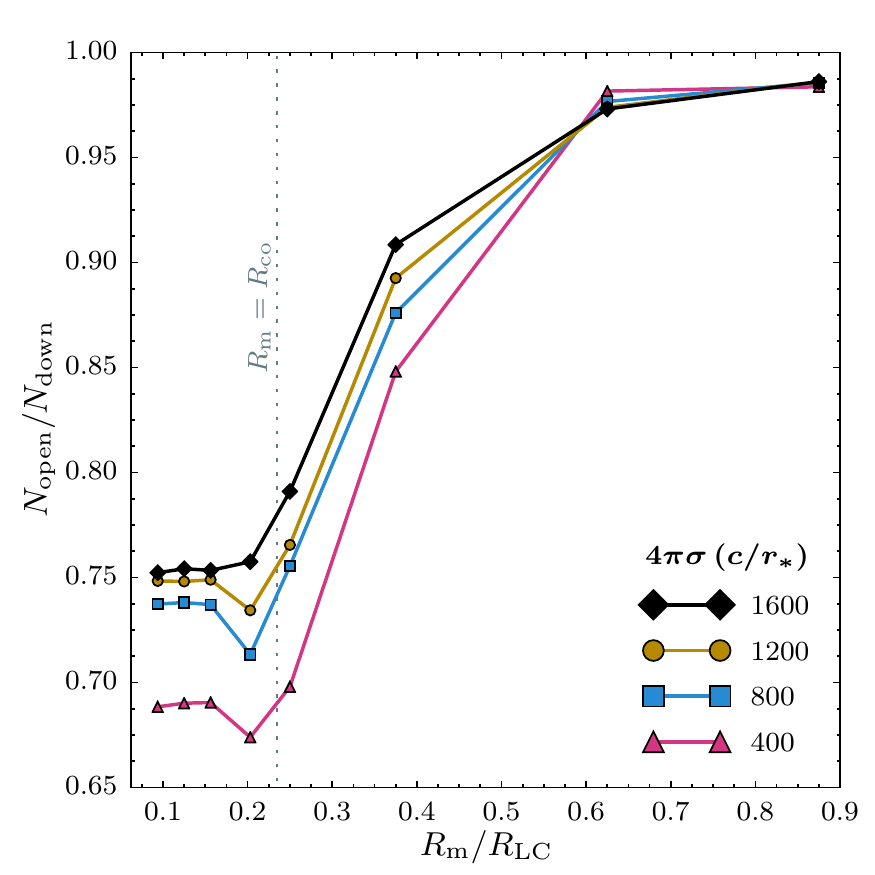}
  \end{center}
  \caption{ \label{fig:torqueOpen} Ratio of spin-down torque on open field lines $\Nopen$ to total spin-down torque $\Ndown$ for the grid of constant-conductivity thin disc simulations.}
\end{figure}

The total spin-down torque on the pulsar $\Ndown$ is shown in Fig.~\ref{fig:torque}, together with the equivalent models, found by relating the open-flux spin-down torque to the open flux via $\Nopen \propto \ffopen^2$. Here the models provide an even better fit to the data, because the small amount of stellar flux closed beyond $\rco$ provides enough spin-down torque to make up for the shortfall in opening efficiency. This is possible because a given field line supplies less torque when open than when closed (except when the closed field line enters the disc very close to $\rco$). For these disc parameters, $\Ndown \approx (\ff_{\rm 0, m}/\ff_{\rm 0, open})^2 \No$ appears to be a good model for the total spin-down torque on the pulsar. Most of the angular momentum is extracted by the open field lines, and the ratio of open-flux torque $\Nopen$ to total spin-down torque $\Ndown$ increases with increasing disc conductivity; see Fig.~\ref{fig:torqueOpen}. 
In general, one might expect that this model will provide an over-estimate of $\Nopen$, because in reality all of the flux outside $\rmag$ will not be opened ($\zeta$ will be less than unity), and an underestimate of $\Ndown$, because Fig.~\ref{fig:torque} shows that $\Ndown$ increases with disc conductivity, and even at the unrealistically low conductivities of our simulations the total spin-down torque already slightly exceeds the model in many cases.

\begin{figure}
  \begin{center}
    \includegraphics[width=3.35in, trim = 1.0mm 1.0mm 1.0mm 2.0mm, clip]{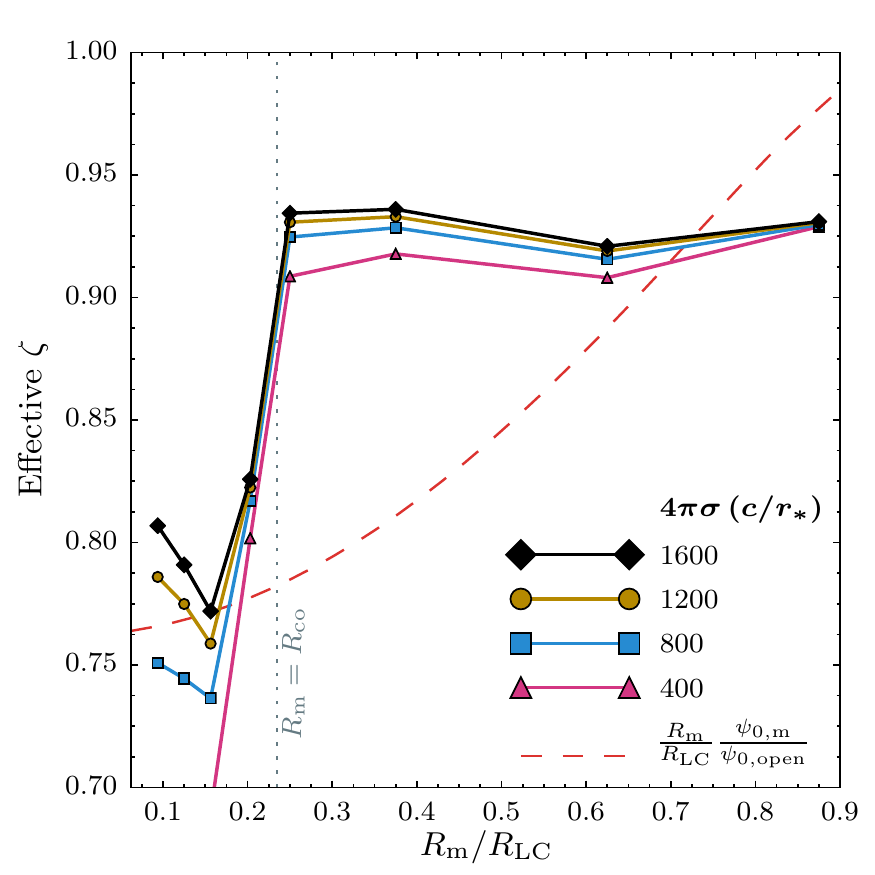}
  \end{center}
  \caption{ \label{fig:zeta} Effective $\zeta$ parameter implied by equating the torques predicted by the model of equation~(\ref{eq:torqueModel}) to the total spin-down torques of Fig.~\ref{fig:torque}. The red dashed line indicates the effective $\zeta$ produced by complete opening of the isolated pulsar's closed zone outside $\rmag$ , i.e.\ the same model as the equivalent lines in Figs.~\ref{fig:flux} and \ref{fig:torque}; the black dashed lines from these figures here corresponds to $\zeta = 1$. }
\end{figure}

Finally, the stellar-torque and jet-power models of equation~(\ref{eq:torqueModel}) are parameterized by $\zeta$, e.g.\ $N = \zeta^2 (\rlc/\rmag)^2 \No$. If we associate this torque $N$ with the total spin-down torque on the star $\Ndown$, we can derive from the simulation data in Fig.~\ref{fig:torque} an effective $\zeta = (\rmag/\rlc) \sqrt{\Ndown/\No}$; this quantity is plotted in Fig.~\ref{fig:zeta}. The model curve (red dashed line) in this figure is the effective $\zeta$ implied by complete opening of the flux in the closed zone beyond $\rmag$ in the isolated pulsar case; note that not all of the spin-down above this line comes from closed flux in the disc, since the opening of field lines changes the magnetic pressure distribution and `pulls out' the underlying magnetic field [compare e.g.\ the innermost field lines drawn in Fig.~\ref{fig:multipanelTime}(a) and (f)].

The $\zeta$ shown in Fig.~\ref{fig:zeta} is useful for describing the \textit{total} spin-down torque in the simulations in terms of the simple model. The jet power, on the other hand, only comprises the energy extracted by open field lines, and so the effective $\zeta$ for jet launching is found by multiplying the above by $\sqrt{\Nopen/\Ndown}$; e.g.\ when $\rmag < \rco$ the effective jet-$\zeta$ for our highest conductivity simulations is lower than indicated in Fig.~\ref{fig:zeta} by a factor of about 0.87.


\section{Comparison to Non-Relativistic Simulations}
\label{sec:comparison}

The results of non-relativistic simulations can be scaled to astrophysical systems of a chosen size by multiplying values in code units by appropriate reference values of the correct dimension; results reported in physical units can be scaled to a different system using the ratio of the corresponding reference values \citep[see e.g.][]{Romanova:2002aa, Zanni:2013aa}. For example, using as basic scales the stellar radius $r_*$, mass $M_*$, and magnetic field strength $B_*$, one can construct a reference velocity via the Keplerian velocity at the stellar surface, $v_{\rm K*} = \sqrt{G M_*/r_*}$, and a density using $\rho_* = B_*^2/v_{\rm K*}^2$. With these scales, reference values for jet power and torque can be composed as $L_* = B_*^2 r_*^2 v_{\rm K*}$ and $N_* = B_*^2 r_*^3$ respectively. 

Below we compare the results of three prior sets of simulations to corresponding values from the basic models of equations~(\ref{eq:torqueModel}) and (\ref{eq:jetModel}) for open-flux spin-down torque and jet power, using values of $\zeta$ from our simulations. We have attempted to match the ratio $\rmag/\rco$ in the models to those reported for the simulations; in other words the comparison is made at the same ``fastness parameter'', $\omega_{\rm fast} = \Omega_*/\Omega_{\rm K}(\rmag) = (\rmag/\rco)^{3/2}$. In all cases we take the neutron star's mass and radius to be $1.4 \msun$ and 10 km.

\subsection{ \citet{Goodson:1999aa} }

The authors describe a jet along the rotation axis, whose matter is supplied by the accretion disc when the disc-star-coupling field lines are roughly dipolar. Differential rotation between star and disc causes the field lines to inflate, as in Section~\ref{sec:inflation}, driving an outflow on the opening field lines between the axis and the forming current layer. Reconnection across the current layer closes the field lines, and begins a new jet cycle. For a star having $r_* = 1.5$~R$_{\sun}$, $M_* = \msun$, and $B_* = 10^3 \G$, the authors report a jet velocity of 140~km~s$^{-1}$ and a excretion rate of $7\times10^{-9} \msunyr$, giving a jet power of 8.7$\times 10^{31} \ergsec$. Scaled to a neutron star having $B_* = 10^8 \G$ this becomes 3.0$\times 10^{34} \ergsec$.

To make a corresponding estimate using equation~(\ref{eq:jetModel}), we see that in the simulation of \citet{Goodson:1999aa} the disc approached to as close as 1.5 $r_*$, which we take as the magnetospheric radius; their stellar spin period of 1.8 days then gives $\rmag/\rco = 0.36$. Comparing to a star rotating at $\nu = 300 \Hz$, as in our simulations, therefore requires $\rmag = 1.36\,r_*$ in the model, giving, using equation~(\ref{eq:jetModel}) and $\zeta = 0.7$, an estimated jet power of 3.2$\times 10^{35} \ergsec$, roughly an order of magnitude larger than the above hydrodynamic energy flux.

\subsection{ \citet{Ustyugova:2006aa} }

In these simulations, in the strong propeller regime (i.e., where the disc is truncated signficantly outside the corotation radius), a large spin-down torque is applied to the star via its magnetic field. The authors estimate a total spin-down torque of $3.9\times 10^{33} \gcmsq$, for a neutron star having $B_* = 10^9 \G$ and $\nu = 795 \Hz$. The disc truncation radius is highly dynamic in their simulation, and we estimate it to be approximately at 3 $\rco$ on average. At 795~Hz, $\rco \approx 2\, r_*$ and $\rlc \approx 6\, r_*$, so holding the fastness parameter constant places the magnetospheric radius roughly at the light cylinder, and so in our model the spin-down torque would be just the torque on the isolated pulsar, $N \approx |N_0| = 4.6 \times 10^{33} \gcmsq$. 

For a less extreme example, one could take a star with a weaker field, $B_* = 10^8 \G$, reducing the scaled torque in the MHD simulation to $3.9\times 10^{31} \gcmsq$. If the stellar spin frequency is 300~Hz, $\rmag/\rco = 3$, and $\zeta = 0.9$ (as this is the total torque on the star), one would estimate a torque, using equation~(\ref{eq:torqueModel}), of $4.1 \times 10^{30} \gcmsq$. In the strong propeller regime, when the accretion flow is truncated relatively near the light cylinder and the open-flux spin-down torque is not significantly increased above its isolated-pulsar value, one may in many cases expect the usual propeller mechanism \citep[e.g.][]{Illarionov:1975aa} to be the strongest effect.

\subsection{ \citet{Zanni:2013aa} }

In this work a set of simulations is described which displays the ``magnetospheric ejections'' of \citet{Goodson:1999aa} and Fig.~1(a--c), both in accreting and weak propeller phases. They generally find that the ejections and the stellar wind apply roughly equal torques on the star, so here we compare their stellar wind torque to our open-flux torque, scaled to two scenarios: a weak-field, moderate-spin neutron star ($B_* = 10^8 \G$, $\nu = 300 \Hz$), and a strong-field, rapidly spinning star ($B_* = 10^9 \G$, $\nu = 600 \Hz$).

In the accreting case, they find that the stellar wind applies $N_{\rm sw} \approx 0.4\, J_*$, where $J_* = M_* r_* v_{\rm K*}$ is the reference angular momentum of the star; therefore one recovers physical units via $N_{\rm sw} = 0.4 \times J_* \times N_*$, where $N_* = B_*^2 r_*^3$. In the weak-moderate and strong-rapid scenarios this translates into torques of 2 $\times 10^{32} \gcmsq$ and 4 $\times 10^{34} \gcmsq$ respectively. Their simulation has $\rco = 4.64\, r_*$ and $\rmag \approx 2.7\, r_*$; maintaining $\rmag/\rco$ gives, for $\zeta = 0.7$, estimates of 6.6 $\times 10^{31} \gcmsq$ and 3.3 $\times 10^{34} \gcmsq$ in each case. 

In the propeller simulation, they find $N_{\rm sw} \sim J_*$, and so the torques are approximately 2.5 times larger than quoted above. In this case the disc is truncated very close to corotation, and so the torques estimated in the open-flux model are about 2.6 times smaller than in the accreting case, and roughly an order of magnitude smaller than those in \citet{Zanni:2013aa}.

\vspace{5mm}

\noindent
The above estimates suggest that magnetically dominated open field lines may, in some cases, make a significant contribution to the extraction of energy and angular momentum from the star. The relativistic wind on the open flux becomes increasingly important as both stellar field strength and rotation rate increase. Effects due to star-disc magnetic coupling can be expected to dominate when the accretion flow is truncted close to the light cylinder and there is little opportunity for flux opening.  However, if the accreting plasma is prevented by the pulsar wind from entering the light cylinder the torque will be almost unchanged from that on the isolated pulsar; the light cylinder is an Alfv\'{e}n surface, across which inward communication via field line twisting or dragging is impossible.


\section{Discussion and Conclusions}
\label{sec:conclusions}

We have performed a suite of simulations in which a relativistic force-free pulsar magnetosphere is coupled to a matter-dominated disc with a fixed velocity field, and finite thickness and conductivity. This approach allows us to study the pulsar--disc corona in the highly magnetically dominated limit, and with very low intrinsic numerical dissipation. We have used two prescriptions for the disc properties: (1) a disc model in which the accretion velocity, conductivity, and disc thickness are self-consistent with a specified Shakura-Sunyaev $\alpha$ parameter, and (2) a simpler thin-disc configuration with constant conductivity. 

We have found two classes of solutions: (1) those in which no true steady state is reached, and where opening and reconnection of the pulsar's magnetic field continues indefinitely, and (2) those which settle eventually into a steady state. Whether a steady state is reached can be determined by the disc's conductivity (more diffusive discs settle more easily into equilibrium), and can be dependent on the presence of the radial accretion velocity, at least for thicker discs where this velocity is larger.

The torque on the star, mediated by its magnetic field and caused both by its rotation and its interaction with the accretion flow, is a useful diagnostic. We find that increasing the disc conductivity causes more of the star's magnetic field to open to infinity, increasing the spin-down torque on the star, and decreasing the spin-up torque. Increasing the disc's thickness at constant $\alpha$ parameter increases the spin-down contribution, and appears to have little effect on the spin-up torque.

Field line opening has long been assumed to be efficient for cold, thin, high-condutivity discs. Here we show that, at least in our idealized simulations, most of the star--disc coupling flux opens also in the case of moderately slim, fairly diffusive discs (i.e. $\alpha \gtrsim 0.4$), and for thick discs.

The spin-down torque on open field lines is only weakly dependent on their poloidal arrangement, whether they are collimated by the disc's magnetic field into a jet or have relaxed into a nearly radial geometry. In our simulations the disc's open magnetic field is gradually expelled outwards, since the ratio of the accretion velocity to the resistive drift due to magnetic annihilation in the disc is $v_{\rm accrete}/v_{\rm resist} \sim h/r$ (see equation~\ref{eq:expulsionSpeed}). The insensitivity of the spin-down torque to the flux distribution in the poloidal plane allows a simple, widely applicable, estimate to be made via the close correspondance between the relaxed, nearly monopolar state and the isolated pulsar magnetosphere. For a given spin frequency, the open-flux spin-down torque is determined almost solely by the number of open field lines, allowing it to be estimated as in equation~(\ref{eq:torqueModel}).

This relationship was used in Paper I to construct a simple torque model for millisecond pulsars surrounded by discs. If one assumes that flux opening is efficient, and therefore that (1) all of the spin-down torque is applied by open field lines, and (2) the spin-up torque is dominated by the accretion torque, $\Nacc \approx \mdot \sqrt{GM\rmag}$, one can estimate the total torque on the star as

\begin{equation}
\Ntot = 
  \begin{dcases}
  \mdot\sqrt{GM\rmag} - \zeta^2\frac{\mu^2}{\rmag^2}\frac{\Omega}{c}, & \rmag < \rco \\
  - \zeta^2\frac{\mu^2}{\rmag^2}\frac{\Omega}{c}, & \rco < \rmag < \rlc\\
  - \mu^2 \frac{\Omega^3}{c^3}, & \rmag > \rlc.
  \end{dcases}
\label{eq:ntot}
\end{equation}
Note that the $\Nacc$ spin-up torque is necessarily absent from our simulations, which do not include the accretion funnel inside $\rmag$. The spin-up torque we measure is that due to field lines coupling the star to the Keplerian disc between $\rmag$ and $\rco$. 

Looking beyond stellar torques, one can estimate the star's potential jet power by associating the electromagnetic pulsar wind with an observable radio jet, as in equation~(\ref{eq:jetModel}). For the enhanced pulsar wind to form a jet, it would need to be collimated by an external source of pressure, for example a disc wind. We do not describe separate diagnostics of the jet power in our simulations, as the angular momentum and energy fluxes are here simply related by a factor of $\Omega$.

We find that the torque and jet power models of equations~(\ref{eq:torqueModel}) and (\ref{eq:jetModel}) provide a good description of our simulations, and that the agreement improves with increasing disc conductivity. In the simulations, the spin-down torque is applied primarily by open field lines, whose share of the total spin-down torque also increases as conductivity increases.

In Paper I, we argued that in the case of efficient flux opening, $\zeta \sim 1$, the enhanced pulsar wind could dominate the spin-down torque on rapidly rotating accreting pulsars, and that the power extracted by the increased open flux would be sufficient to account for the relativistic jets from neutron-star X-ray binaries. We find that the opening efficiency in the simulations is indeed close to unity, even when the disc is more diffusive than realistically expected. In our constant-conductivity thin-disc runs, we find the effective $\zeta$ to be $\gtrsim 0.8$ when the disc extends inside the corotation radius, and $\gtrsim 0.9$ if it is truncated beyond corotation. In our $\alpha$-disc reference simulations, the $\zeta$ parameter is $\sim 0.95$ for run R1 ($\alpha = 0.1$), and $\sim 0.8$ for R2 ($\alpha = 0.4$) and R3 ($\alpha = 0.1$, thicker disc). These values are found via $\zeta = (\rmag/\rlc) \sqrt{\Ndown/\No}$ where $\Ndown$ is the total spin-down torque on the star; they suggest that our open-flux model may provide reasonable first estimates for torques and jet powers for those systems whose accretion flows are sufficiently similar to those investigated here. 

Our simulations indicate that, in the quasi-steady state, the frequency and amplitude of the torque oscillations depend on the properties of the disc, when the stellar rotation rate and the disc's inner radius remain constant. For example, Fig.~\ref{fig:torqueOverTime} shows that the oscillations occur at higher frequency in the case of the thicker disc, at the same $\alpha$ parameter. While short-timescale variations in the derivative of the spin frequency may be very challenging to detect during accretion episodes, the high-energy emission from the magnetic reconnection which accompanies the oscillations may leave an imprint on the system's X-ray variability spectrum. We defer an exploration of this effect to future work.

In the simulations described here the disc's location is fixed and the initial magnetic field is a potential dipole frozen into the star. We have also performed simulations in which an isolated pulsar magnetosphere is first formed, and the disc is then moved inwards from beyond the light cylinder. For the ranges of parameters discussed above, provided the disc's effective magnetic Prandtl number $\Prm$ is near unity, the same steady or quasi-steady states are eventually reached, once the inwardly pinched field lines have diffused back through the disc. If the accretion velocity is larger than the outward resistive-annihilation velocity (i.e.\ $\Prm$ is greater than $\sim r/h$; see Fig.~\ref{fig:multipanelPrm100}) the inwardly pinched field lines may be trapped in the inner part of the disc; solutions of this kind will be further investigated in a future paper. 

There are several important aspects of the pulsar--disc problem which are absent from our simulations. The disc has been included in a simplified manner rather than self-consistently, and so, for example, the dependence of the accretion rate on the disc's local large-scale magnetic field is not captured. The simulations are axisymmetric, and therefore lack the azimuthally dependent field opening resulting from obliquity between the spin and magnetic axes. Obliquity can be expected to modify our torque and jet power estimates, since a given disc profile will intersect a different bundle of field lines as the obliquity is varied. We have assumed a corona with vanishing inertia, but as the disc becomes thicker the density throughout the corona should increase, which may impede field line opening but cause more torque to be applied per open field line. Given these concerns, care must be taken when interpreting observations in the light of these results.

\vspace{0.2cm}
KP was supported by the Max Planck-Princeton Center for Plasma Physics, and by NASA through Einstein Postdoctoral Fellowship grant number PF5-160142 awarded by the Chandra X-ray Center, which is operated by the Smithsonian Astrophysical Observatory for NASA under contract NAS8-03060. AS acknowledges support from NASA grants NNX14AQ67G, NNX15AM30G, and a Simons Investigator Award from the Simons Foundation. AMB acknowledges support from NASA grant NNX13AI34G and a Simons Investigator Award from the Simons Foundation.

The simulations presented in this article were performed on computational resources supported by the Princeton Institute for Computational Science and Engineering (PICSciE) and the Office of Information Technology's High Performance Computing Center and Visualization Laboratory at Princeton University, and on the SAVIO computational cluster resource provided by the Berkeley Research Computing program at the University of California, Berkeley (supported by the UC Berkeley Chancellor, Vice Chancellor of Research, and Office of the CIO).



\end{document}